\documentclass{article}

\usepackage{preprint}

\usepackage[pdftex]{graphicx}
\usepackage[utf8]{inputenc}
\usepackage[T1]{fontenc}
\usepackage{hyperref}
\usepackage{url}
\usepackage{booktabs}
\usepackage{amsfonts}
\usepackage{nicefrac}
\usepackage{microtype}
\usepackage{lipsum}
\usepackage{multicol}
\usepackage{enumitem}
\setlist{itemsep=0pt}
\setlist{topsep=1pt}

\usepackage{booktabs}
\usepackage{multirow}
\usepackage{stmaryrd}
\usepackage{algorithm}
\usepackage{algpseudocode}
\usepackage{bm}
\usepackage{amsmath,amssymb,amsthm}
\usepackage[export]{adjustbox}
\newtheorem{lemma}{Lemma}
\newtheorem{proposition}{Proposition}

\newtheorem{theorem}{Theorem}

\usepackage{siunitx}

\usepackage{tikz}
\usepackage{fig/tikz-style}
\usepackage[font=small,subrefformat=parens]{subcaption}
\usepackage{macro}

\title{Priority Inheritance with Backtracking\\for Iterative Multi-agent Path Finding}
\author{
  Keisuke Okumura\\
  Tokyo Institute of Technology\\
  \texttt{okumura.k@coord.c.titech.ac.jp}
  \And
  Manao Machida\\
  NEC Corporation\\
  \texttt{manaomachida@nec.com}
  \And
  Xavier D\'{e}fago\\
  Tokyo Institute of Technology\\
  \texttt{defago@c.titech.ac.jp}
  \And
  Yasumasa Tamura\\
  Tokyo Institute of Technology\\
  \texttt{tamura@c.titech.ac.jp}
}

\begin{document}
\maketitle
\begin{abstract}
  In the Multi-Agent Path Finding (MAPF) problem, a set of agents moving on a graph must reach their own respective destinations without inter-agent collisions. In practical MAPF applications such as navigation in automated warehouses, where occasionally there are hundreds or more agents, MAPF must be solved iteratively online on a lifelong basis. Such scenarios rule out simple adaptations of offline compute-intensive optimal approaches; and scalable sub-optimal algorithms are hence appealing for such settings. Ideal algorithms are scalable, applicable to iterative scenarios, and output plausible solutions in predictable computation time.

  For the aforementioned purpose, this study presents Priority Inheritance with Backtracking (PIBT), a novel sub-optimal algorithm to solve MAPF iteratively. PIBT relies on an adaptive prioritization scheme to focus on the adjacent movements of multiple agents; hence it can be applied to several domains. We prove that, regardless of their number, all agents are guaranteed to reach their destination within finite time when the environment is a graph such that all pairs of adjacent nodes belong to a simple cycle (e.g., biconnected). Experimental results covering various scenarios, including a demonstration with real robots, reveal the benefits of the proposed method. Even with hundreds of agents, PIBT yields acceptable solutions almost immediately and can solve large instances that other established MAPF methods cannot. In addition, PIBT outperforms an existing approach on an iterative scenario of conveying packages in an automated warehouse in both runtime and solution quality.
\end{abstract}

\section{Introduction}
In systems with multiple moving agents, valid paths must be provided to ensure that agents reach their destinations smoothly without any collisions, while minimizing excess travel time. This problem is embodied in \emph{Multi-Agent Path Finding (MAPF)}~\cite{stern2019definition}, which aims to find a set of collision-free paths on graphs. Currently,
MAPF is receiving considerable attention because it is highly applicable in problems such as intersection management~\cite{dresner2008multiagent}, automated warehouse~\cite{wurman2008coordinating}, games~\cite{silver2005cooperative}, airport surface operation~\cite{morris2016planning}, and parking~\cite{okoso2019multi}.
However, the optimization of this problem is known to be computationally intractable ~\cite{yu2013structure,ma2016multi,yu2015intractability,banfi2017intractability} because of the exponential growth of the search space as the number of agents increases.

Prior research on MAPF primarily focuses on solving a “one-shot” version of the problem, where agents reach their respective goals from their initial positions only once. In practical applications such as conveying packages in a warehouse with hundreds or more robots~\cite{wurman2008coordinating}, MAPF must be solved \emph{online} and \emph{iteratively} on a \emph{lifelong} basis. That is, whenever an agent reaches a goal, it receives a new one, and the plan, a set of paths, is updated in real-time. Such scenarios rule out simple adaptations of offline and compute-intensive optimal approaches because, despite state-of-the-art optimal algorithms~\cite{lam2020new,li2021pairwise}, it is challenging to find solutions for a few hundred agents within a realistic timeframe. For this purpose, an attractive possibility is to design scalable sub-optimal algorithms that output plausible solutions within a predictable computational time, which also works in iterative situations. Such algorithms are also promising for real-time planning where deliberation time is limited because it enables iterative refinement for known MAPF solutions~\cite{okumura2021iterative}.

To this end, a novel sub-optimal algorithm named \emph{Priority Inheritance with Backtracking (PIBT)} is proposed to solve MAPF iteratively. Theoretically, PIBT ensures \emph{reachability}, that is, every agent always reaches its destination within a finite time when the environment is a graph such that all pairs of adjacent nodes belong to a simple cycle (e.g., biconnected).
PIBT also has low-cost time complexity. It repeats one-timestep planning until termination (e.g., until all agents have reached their destinations). For each timestep, PIBT requires the time complexity $O(|A|\cdot(\Delta(G) + F + \log{|A|}))$, where $|A|$ is the number of agents, $\Delta(G)$ is a diameter of a graph $G$, and $F$ is the time required to evaluate the distance from one vertex to the destination. Note that $F$ can be constant with preprocessing. As a result, PIBT is expected to work in a short runtime, even in the case of massive problems.

From the empirical end, PIBT immediately returns acceptable solutions despite hundreds or more agents in running times that are orders of magnitudes faster than other established MAPF approaches, such as sub-optimal priority-based~\cite{silver2005cooperative}, rule-based~\cite{luna2011push}, and search-based~\cite{li2021eecbs} algorithms, as well as state-of-the-art optimal search-based~\cite{sharon2015conflict} and compiling-based~\cite{lam2019branch} algorithms.
For instance, using an ordinary laptop, we observe that PIBT solves one-shot MAPF in a large grid map ($530\times 481$, number of vertices: $43,151$) from~\cite{stern2019definition} with $1,000$ agents in at most 5 seconds, while keeping sub-optimality below $1.5$ on average.
Furthermore, PIBT outperforms an existing approach for both runtime and solution quality on an iterative scenario of conveying packages in an automated warehouse, called Multi-Agent Pickup and Delivery (MAPD)~\cite{ma2017lifelong}.

Reachability does not ensure that all agents are on their goals simultaneously; hence, PIBT is incomplete for conventional MAPF. However, agents often do not need to stay at their goals in real-life applications such as lifelong delivery tasks. Reachability is convenient in such scenarios. Indeed, this study introduces the application of PIBT to a lifelong variant of MAPF (i.e., MAPD~\cite{ma2017lifelong}) while ensuring completeness.

The mechanism of PIBT is simple and easy to implement. The algorithm focuses on the adjacent movements of multiple agents based on prioritized planning~\cite{erdmann1987multiple} in a unit-length time window. Priority inheritance is a well-known approach for dealing effectively with priority inversion in real-time systems~\cite{sha1990priority} and is applied here to path adjustment. When a low-priority agent--X impedes the movement of a higher-priority agent--Y, agent--X temporarily inherits the higher-priority of agent--Y. To avoid agents getting stuck waiting, priority inheritance is executed in combination with a backtracking protocol. Because this mechanism only requires local interactions between agents, PIBT has a high potential for decentralized implementations.

In summary, the main contributions are two-fold:
\begin{enumerate}
\item We propose the PIBT algorithm, which solves MAPF iteratively and guarantees reachability.
\item We evaluate the algorithm in various environments to assess its practicality.
  In particular, experimental results analyzing various scenarios, including robot demonstration, confirm the adequacy of the algorithm both in finding paths in large environments with many agents as well as in conveying packages in an automated warehouse.
\end{enumerate}
All external materials (e.g., code and a movie) are available at \url{https://kei18.github.io/pibt2}. The techniques presented here provide practical insight into online and lifelong scenarios involving a swarm of agents.

The rest of the paper is organized as follows.
Section~\ref{sec:relatedwork} reviews existing MAPF studies.
Section~\ref{sec:pre} defines MAPF and its variant called MAPD, a problem considering operations in automated warehouses.
Then, Section~\ref{sec:pibt} presents the PIBT algorithm along with a theoretical analysis of its reachability and time complexity.
In Section~\ref{sec:eval}, we empirically evaluate PIBT using many simulated scenarios and a demonstration on real robots.
Finally, Section~\ref{sec:conclusion} concludes the paper and discusses future directions.

\paragraph{Difference from Conference Versions}
The PIBT algorithm was initially presented in our preliminary paper~\cite{okumura2019priority}.
PIBT$^+$ (Sec.~\ref{sec:pibt-mapf}) was initially mentioned in~\cite{okumura2021iterative}; however, theoretical analyses were not provided; we remedy this shortfall here. This paper, in addition to improving the description and presentation of PIBT, has the following major differences from the previous versions:
(1)~Exhaustive experiments were conducted on one-shot MAPF using various fields and solvers.
As a result, this paper thoroughly clarifies the advantages and disadvantages of PIBT compared to other de facto approaches. The implementation was also significantly improved, with the running time decreased by orders-of-magnitudes than that of the conference version~\cite{okumura2019priority}. This is mainly due to distance evaluation, as explained in Section~\ref{sec:pibt-mapf}.
(2)~Further theoretical analyses were conducted on PIBT, for instance, on reachability considering rotation conflicts.
(3)~A stress test for the number of agents and the robot demonstration of PIBT in an iterative scenario were added.

\section{Related Work}
\label{sec:relatedwork}
This section describes the basis of MAPF, summarizes the relationship between PIBT and other existing MAPF algorithms, and presents lifelong and online situations apart from conventional MAPF studies.

\subsection{Basis of Multi-Agent Path Finding (MAPF)}
The \emph{MAPF}~\cite{stern2019definition} assigns each agent on a graph a collision-free path to its destination. Time is discrete;  for each timestep, each agent performs atomic action in a synchronized manner such that it either moves to an adjacent node or stays in its current location.
MAPF is known to be an NP-hard problem according to various optimization criteria~\cite{surynek2010optimization,yu2013structure}, which holds for
restricting fields in planar graphs~\cite{yu2015intractability}, in grid structures~\cite{banfi2017intractability}, and in approximating within any constant factor less than 4/3 for the last arrival time metric~\cite{ma2016multi}. Limiting the discussion to sub-optimal solutions on undirected graphs, there exists a procedure of $O(|V|^3)$ operations~\cite{kornhauser1984coordinating} for the compatibility of the pebble motion problem (a generalization of the sliding tile puzzle), where $|V|$ is the number of vertices. In contrast, finding solutions on directed graphs itself is NP-complete~\cite{nebel2020computational}. The case of strongly biconnected digraphs is special, where there is a polynomial-time procedure~\cite{botea2018solving} to solve MAPF for such graphs. \emph{The proposed method, PIBT, is neither complete nor optimal for MAPF. However, PIBT works with both directed and undirected graphs.}

\subsection{Algorithms}
To date, the numerous solvers developed for MAPF can be categorized into optimal or sub-optimal, complete or incomplete.
Here, an algorithm is \emph{complete} when it ensures to return a set of collision-free paths such that the end of each path is the destination of each agent.
If no such set of paths exists, complete algorithms must report its absence.
Another categorization is solving style, by which algorithms are classified into the following five classes:
\begin{itemize}
\item \emph{Search-based} approaches search feasible solutions in a coupled manner among agents.
  Examples are \astar with operator decomposition~\cite{standley2010finding}, increasing cost tree search~\cite{sharon2013increasing}, enhanced partial expansion \astar~\cite{goldenberg2014enhanced}, conflict-based search (CBS)~\cite{sharon2015conflict}, and M$^\ast$~\cite{wagner2015subdimensional}.
  These algorithms are complete and can solve MAPF optimally.
  Bounded sub-optimal versions~\cite{barer2014suboptimal,li2021eecbs}, which are complete, have also been developed to solve large instances.
\item \emph{Compiling-based} approaches reduce MAPF to well-known problems, such as, SAT~\cite{surynek2012towards,surynek2016efficient,surynek2019unifying}, integer linear programming~\cite{yu2016optimal,lam2019branch}, and answer set programming~\cite{erdem2013general,gomez2020solving}.
  Those algorithms are complete and optimal.
\item \textbf{\emph{Prioritized planning}}~\cite{erdmann1987multiple} approaches sequentially plan individual paths for agents in a decoupled manner; in general, these algorithms are neither complete nor optimal.
\item \textbf{\emph{Rule-based}} approaches make agents move step-by-step following ad-hoc rules. Examples include a graph abstraction approach~\cite{ryan2008exploiting}, \cite{peasgood2008complete} for spanning trees, BIBOX~\cite{surynek2009novel} for biconnected graphs, and push and swap/rotate~\cite{luna2011push,de2013push} for arbitrary graphs.
  The rule-based approaches are sub-optimal but often ensure completeness for certain problem instances.
\item \emph{Learning-based} approaches use machine learning techniques and imitate good behavior of agents for expert data.
  Examples include reinforcement learning~\cite{sartoretti2019primal,damani2021primal} and graph neural networks~\cite{li2020graph,li2021message}.
  They are suboptimal and incomplete.
\end{itemize}
See~\cite{felner2017search,stern2019multi} for comprehensive reviews.
Despite state-of-the-art search-based or compiling-based optimal approaches, planning with a few hundred agents within a reasonable time is challenging~\cite{li2021pairwise,lam2020new}. Learning-based approaches scale well in certain cases; however, they do not have any theoretical guarantees. Therefore, for the target scenarios considered in this study (i.e., online and lifelong scenarios with hundreds or more agents), prioritized planning and rule-based approaches are promising methods. Indeed, PIBT combines these two classes. The following reviews focus on prioritized planning and rule-based approaches. Note that PIBT is ``incomplete'' because it ensures that all agents eventually reach their destinations; however, they may not be there simultaneously. This property is designed for lifelong situations and significantly differs from the theoretical properties provided by existing algorithms for one-shot scenarios.

Prioritized planning~\cite{erdmann1987multiple} is neither complete nor optimal; however, it is a computationally inexpensive approach to MAPF. It is based on the concept that agents sequentially plan paths according to their unique priorities while avoiding conflicts with previously planned paths. Because of its computational efficiency and simplicity, prioritized planning is highly favorable in practice and is often used as part of MAPF solvers. For instance, Wang and Botea~\cite{wang2011mapp} combined techniques of prioritized planning for the way the blanks moved around in sliding tile puzzles and proposed a complete algorithm for a specific class of instances. Velagapudi \etal~\cite{velagapudi2010decentralized} proposed decentralized implementations of prioritized planning that divide planning into several iterations. Variants of prioritized planning have also been studied, e.g., multi-robot decoupled planning that first computes individual plans neglecting collisions, and then replans individual paths while incorporating collision costs gradually~\cite{jiang2019multi}. We acknowledge that this scheme has similarities to PIBT, albeit the applied problem is different and the approach has no theoretical guarantee. The well-known algorithm of prioritized planning for MAPF is hierarchical cooperative \astar (\hca)~\cite{silver2005cooperative}. The “windowed” version of \hca is known as windowed \hca (\whca)~\cite{silver2005cooperative}, which repeatedly plans fixed-length (windowed) paths for all agents. PIBT can be considered as using \whca with a unit-length window sizes.

Because priority ordering is crucial for prioritized planning, several studies addressed this issue. The negotiation process between agents regarding the ordering was studied in~\cite{azarm1997conflict}; this process solves conflicts by having involved agents try all priority orderings and deals with congestion by limiting negotiation to at most three agents while letting others wait. A similar approach is taken in~\cite{bnaya2014conflict} where the winner determines the prioritization scheme. There are some heuristic approaches for offline planning to find a reasonable ordering, for instance, adjusting the ordering by simple hill-climbing~\cite{bennewitz2002finding}, or using distances between initial locations and destinations~\cite{van2005prioritized}. There is a sufficient condition that the sequential collision-free solution can always be constructed regardless of priority orders, called well-formed instances~\cite{vcap2015prioritized}, such that for each pair of start and goal, a path exists that traverses no other starts and goals. However, well-formed instances are difficult to realize in dense situations. A recent theoretical analysis for prioritized planning~\cite{ma2019searching} identifies instances that fail for any order of \emph{static} priorities, which provides a strong case for planning based on \emph{dynamic} priorities, such as the approach taken with PIBT. The paper also presents a variant of prioritized planning that searches for good static priorities using techniques from CBS; however, the method has no theoretical guarantee.

Push and swap/rotate (PS)~\cite{luna2011push,de2013push}, which partly influenced the proposed PIBT, is a sub-optimal rule-based approach for arbitrary graphs. It relies on two primitives: the ``push'' operation to move an agent toward its goal and the ``swap'' operation to allow two agents to swap locations without altering the configuration of other agents. These approaches, however, only allow a single agent or a pair of agents to move at a time. Some studies enhanced PS, for instance, parallel push and swap~\cite{sajid2012multi} by enabling all agents to move simultaneously, or push--swap--wait~\cite{wiktor2014decentralized} by taking a decentralized approach in narrow passages. Wei \etal~\cite{wei2014multi} proposed a decentralized approach that partly uses the push and swap technique to avoid deadlocks. DisCoF~\cite{zhang2016discof}, a decentralized method for MAPF, also uses swap operation to ensure completeness. There is an algorithm similar to PS, called tree-based agent swapping strategy (TASS)~\cite{khorshid2011polynomial}, targeting trees. PIBT can be regarded as a combination of safe ``push'' operations because of the backtracking protocol and dynamic priorities. Note that PIBT does not require the operational equivalent of ``swap.''

\subsection{Lifelong and Online Situations}
Conventional MAPF focuses on solving a \emph{``one-shot''}%
\footnote{%
    A \emph{``one-shot''} problem is a problem that has clearly defined inputs and must be solved only once. This is in contrast to problems that must be solved repeatedly for a \emph{stream} of inputs.
}
and \emph{offline} version of the problem, that is, ensuring that agents reach their goal from their initial position just once.
Several variants of the MAPF problem~\cite{ma2017lifelong,vsvancara2019online,li2020lifelong} aim to address the needs of more realistic scenarios, including lifelong and online operations. Among them, \emph{Multi-Agent Pickup and Delivery (MAPD)}~\cite{ma2017lifelong} abstracts real scenarios such as an automated warehouse and has both task allocation and path planning. Agents are assigned to a task from a stream of delivery tasks and must consecutively visit both a pickup and a delivery location. The study proposes two decoupled algorithms based on \hca for MAPD, called token passing and token passing with task swap, respectively. These algorithms can be decentralized but require a certain amount of non-task endpoints where agents do not block other agents' motions.
\emph{Online MAPF}~\cite{vsvancara2019online} addresses situations with a dynamic group of agents, i.e., new agents might appear at runtime, or agents might disappear when they reach their destinations.
We do not address this problem; however, the adaptation of PIBT for this is straightforward because PIBT essentially repeats one-timestep planning.

\section{Problem Definition}
\label{sec:pre}
This study focuses on two problems: conventional one-shot MAPF problem~\cite{stern2019definition} and one of its variants, the MAPD problem~\cite{ma2017lifelong} for lifelong and online scenarios. We formulate both problems in this section. Note that the adaptation of PIBT to other lifelong MAPF variants~\cite{vsvancara2019online,li2020lifelong} is straightforward.

\subsection{Multi-Agent Path Finding (MAPF)}
\label{sec:def:mapf}
An instance of \emph{MAPF} is given by a directed graph $G = (V, E)$, a set of agents $A=\{a_1, \ldots ,a_n\}$ with $|A| \leq |V|$, an injective initial state function $s: A \mapsto V$, and an injective goal state function $g: A \mapsto V$.
To simplify the notation, let $s_i$ and $g_i$ denote $s(a_i)$ and $g(a_i)$, respectively.
Considering practical situations, a graph~$G$ must satisfy the following two properties:
(1)~\emph{simple}, i.e., neither (self) loops nor multiple edges (between the same vertices) are present.
(2)~\emph{strongly-connected}, i.e., every node is reachable from every other node.
This includes all simple undirected graphs that are connected.

Let $\loc{i}{t} \in V$ denote the location of an agent $a_i$ at discrete time~$t \in \mathbb{N}$.
At each timestep~$t$, $a_i$ can move to an adjacent node, or stay at its current location, i.e., $\loc{i}{t+1} \in \neigh{\loc{i}{t}} \cup \{ \loc{i}{t} \}$, where \neigh{v} is the set of nodes neighboring node~$v$ in graph~$G$.
Agents must avoid two types of conflicts:
(1)~\emph{vertex conflict}: $\exists i, j, \;i \neq j, \;\loc{i}{t} = \loc{j}{t}$ (when agents share the same location), and,
(2)~\emph{swap conflict}: $\exists i, j, \;i \neq j, \;\loc{i}{t} = \loc{j}{t+1} \land \loc{i}{t+1} = \loc{j}{t}$ (when two agents swap their occupying vertices).
We refer to a set of paths as \emph{conflict-free} when those two conditions are satisfied.

A solution to \emph{MAPF} is a set of conflict-free paths $\{\path{1}, \ldots, \path{n}\}$ such that all agents reach their goals at a certain timestep~$T$.
More precisely, the problem assigns a path $\path{i} = (\loc{i}{0}, \loc{i}{1}, \dots, \loc{i}{T})$ to each agent such that $\loc{i}{0} = s_i$ and $\loc{i}{T}=g_i$.
Note that all agents must be at their goals at the same timestep.
Hence, $T$ is common among agents.

Two different metrics are commonly used to evaluate the quality of MAPF solutions:
(1)~\emph{sum-of-costs}: $\sum_{a_i \in A} T_i$ where $T_i$ is the earliest timestep such that $\loc{i}{T_i}=g_i, \dots,\loc{i}{T}=g_i, T_i \leq T$,
(2)~\emph{makespan}, i.e., the value of~$T$.

\subsection{Multi-Agent Pickup and Delivery (MAPD)}
Similar to MAPF, a graph $G=(V, E)$ and a set of agents $A = \{ a_1, \ldots, a_n \}$ with $|A| \leq |V|$, and an injective initial state function $s: A \mapsto V$ (for simplicity, $s(a_i)$ is denoted as $s_i$), are given.

Consider a stream of tasks $\Gamma = \{ \tau_1, \tau_2, \ldots \}$.
A task $\tau_j$ is defined as a tuple $(v_{\textit{pickup}}, v_{\textit{delivery}})$, where $v_{\textit{pickup}}, v_{\textit{delivery}} \in V$.
MAPD includes situations in which tasks are added to $\Gamma$ as time progresses. In other words, it is not assumed that all tasks are known initially. An agent is \emph{free} when it has no assigned task. A task $\tau_j \in \Gamma$ can only be assigned to free agents and is assigned to at most one agent. When $\tau_j$ is assigned to $a_i$, $a_i$ has to visit $v_{\textit{pickup}}$ and $v_{\textit{delivery}}$ in order. When the two nodes have been visited by $a_i$, $\tau_j$ is completed, and $a_i$ becomes free again. A \emph{service time} of $\tau_j$ is defined as the time interval from the generation of $\tau_j$ to its completion. Similar to MAPF, time is discretized. At each timestep, each agent can move to an adjacent node or stay at its current node, provided that no two agents collide, i.e., vertex and swap conflicts are prohibited.

The objective is to complete all tasks as quickly as possible.
A solution to \emph{MAPD} is a set of conflict-free paths and task assignments, with each agent starting from $s_i$. When $|\Gamma|$ is finite, an MAPD algorithm is \emph{complete} as it ensures that all tasks are finished within a finite number of timesteps.

Two kinds of objective functions are considered:
(1)~\emph{service time} and (2)~\emph{makespan}, that is, the timestep when all tasks are completed.
Note that makespan is only defined when $\Gamma$ is finite.

\subsection{Other Notations}
The diameter of graph~$G$ is denoted by $\mathit{diam}(G)$, and its maximum degree by $\Delta(G)$. Let \dist{u}{v} denote the shortest path length from $u \in V$ to $v \in V$.

\section{Priority Inheritance with Backtracking (PIBT)}
\label{sec:pibt}
This section introduces the PIBT algorithm. The concept of PIBT is initially explained by intuitive examples. To avoid unnecessarily complex explanations, PIBT is introduced as a centralized algorithm, focusing on analyzing the prioritization scheme itself. Note that PIBT relies on a decoupled approach, rendering it readily amenable to decentralization, as briefly discussed later. After providing a theoretical analysis on PIBT including reachability, how to apply PIBT to specific problems, such as one-shot MAPF and MAPD, is described.

\subsection{Concept}
Aiming to solve MAPF iteratively, PIBT repeats one-timestep prioritized planning until it is terminated. At every timestep, each agent first updates its unique priority. Then, the agents sequentially determine their next location in decreasing order of priorities while avoiding nodes that have been requested by higher-priority agents. Prioritization alone, however, can still cause deadlocks. As shown in Fig.~\ref{fig:pi:stuck}, a stuck agent $a_1$ cannot go anywhere without collisions with other agents; hence, this situation can be regarded as a certain kind of deadlock.

\begin{figure}[t]
 \centering
 \begin{tabular}{ccc}
   \begin{minipage}[t]{0.29\hsize}
   \centering
   \begin{tikzpicture}
     \node[vertex](v1) at (0, 0) {};
     \node[vertex,right=0cm of v1](v2) {};
     \node[vertex,right=0cm of v2](v3) {};
     \node[vertex,above=0cm of v1](v4) {};
     \node[vertex,right=0cm of v4](v5) {};
     \node[vertex,right=0cm of v5](v6) {};
     \node[agenthigh](a3) at (v6) {$a_3$};
     \node[above=0.15cm of a3]() {\scriptsize priority: \textbf{high}};
     \node[agent](a2) at (v1) {$a_2$};
     \node[below=0.15cm of a2]() {\scriptsize medium};
     \node[agentlow](a1) at (v3) {$a_1$};
     \node[below=0.15cm of a1,color=white]() {\scriptsize \textbf{high}};  
     \node[below=0.15cm of a1]() {\scriptsize low};
     \node[dest] (dest3) at (v3) {};
     \node[smalldest] (dest2) at (v2) {};
     \draw[move](a3)--(dest3);
     \draw[move](a2)--(dest2);
   \end{tikzpicture}
   \subcaption{Stuck agent}
   \label{fig:pi:stuck}
  \end{minipage}
  &
  \begin{minipage}[t]{0.29\hsize}
   \centering
   \begin{tikzpicture}
     \node[vertex](v1) at (0, 0) {};
     \node[vertex,right=0cm of v1](v2) {};
     \node[vertex,right=0cm of v2](v3) {};
     \node[vertex,above=0cm of v1](v4) {};
     \node[vertex,right=0cm of v4](v5) {};
     \node[vertex,right=0cm of v5](v6) {};
     \node[agenthigh](a3) at (v6) {$a_3$};
     \node[above=0.15cm of a3]() {\scriptsize \textbf{high}};
     \node[agent](a2) at (v1) {$a_2$};
     \node[below=0.15cm of a2]() {\scriptsize medium};
     \node[agent](a1) at (v3) {$a_1$};
     \node[below=0.15cm of a1]() {\scriptsize low (as \textbf{high})};
     \node[dest] (dest3) at (v3) {};
     \node[smalldest] (dest1) at (v2) {};
     \draw[move](a3)--(dest3);
     \draw[move](a1)--(dest1);
   \end{tikzpicture}
   \subcaption{Priority inheritance}
   \label{fig:pi:pi}
  \end{minipage}
  &
  \begin{minipage}[t]{0.29\hsize}
   \centering
   \begin{tikzpicture}
     \node[vertex](v1) at (0, 0) {};
     \node[vertex,right=0cm of v1](v2) {};
     \node[vertex,right=0cm of v2](v3) {};
     \node[vertex,above=0cm of v1](v4) {};
     \node[vertex,right=0cm of v4](v5) {};
     \node[vertex,right=0cm of v5](v6) {};
     \node[agenthigh](a3) at (v3) {$a_3$};
     \node[agent](a2) at (v1) {$a_2$};
     \node[agent](a1) at (v2) {$a_1$};
     \node[below=0.15cm of a1,color=white]() {\scriptsize \textbf{high}};  

   \end{tikzpicture}
   \subcaption{One timestep later}
   \label{fig:pi:result}
  \end{minipage}
 \end{tabular}
 \caption{
   \textbf{Examples of priority inheritance.}
   Circles with solid lines represent agents.
   Requests for the next timestep are depicted by dashed circles, determined greedily according to agents’ destinations (omitted here).
   Without inheritance (\ref{fig:pi:stuck}), a stuck agent ($a_1$) cannot give way to a high-priority agent ($a_3$) without risking collision into a third agent ($a_2$).
 With priority inheritance (\ref{fig:pi:pi}), $a_1$ temporarily inherits the priority of $a_3$ forcing $a_2$ to solve the situation (\ref{fig:pi:result}).
 }
 \label{fig:pi}
\end{figure}

\subsubsection{Priority Inheritance}
A situation with a stuck agent can also be regarded as a case of \emph{priority inversion}, in the sense that a low-priority agent ($a_1$) holding a resource claimed by a higher-priority agent ($a_3$) fails to obtain a second resource held by an agent with medium priority $a_2$. A classical way to deal with priority inversion is to rely on \emph{priority inheritance}~\cite{sha1990priority} (Fig.~\ref{fig:pi:pi}). The rationale is that a low-priority agent ($a_1$) temporarily inherits the higher priority of the agent ($a_3$) claiming the resources it holds, thus forcing medium-priority agents ($a_2$) out of the way (Fig.~\ref{fig:pi:result}).

\begin{figure}[t]
 \centering
 \begin{tabular}{cc}
  \begin{minipage}[t]{0.45\hsize}
   \centering
   \begin{tikzpicture}
    \node[vertex](v1) at (0, 0) {};
    \node[vertex,right=0cm of v1](v2) {};
    \node[vertex,right=0cm of v2](v3) {};
    \node[vertex,right=0cm of v3](v4) {};
    \node[vertex,above=0cm of v1](v5) {};
    \node[vertex,right=0cm of v5](v6) {};
    \node[vertex,right=0cm of v6](v7) {};
    \node[vertex,right=0cm of v7](v8) {};
    \node[agenthigh](a7) at (v6) {$a_7$};
    \node[above=0.15cm of a7]() {\scriptsize priority: \textbf{high}};
    \node[agent](a6) at (v7) {$a_6$};
    \node[agent](a5) at (v3) {$a_5$};
    \node[agent](a4) at (v4) {$a_4$};
    \node[agentlow](a3) at (v8) {$a_3$};
    \node[agent](a1) at (v2) {$a_1$};
    \node[below=0.15cm of a1,color=white]() {\scriptsize \textbf{high}}; 
    \node[below=0.15cm of a1]() {\scriptsize low};
    \node[agent](a2) at (v1) {$a_2$};
    \node[below=0.15cm of a2]() {\scriptsize medium};
    \node[dest] (dest7) at (a6) {};
    \node[dest] (dest6) at (a5) {};
    \node[dest] (dest5) at (a4) {};
    \node[dest] (dest4) at (a3) {};
    \draw[move] (a7)--(dest7);
    \draw[move] (a6)--(dest6);
    \draw[move] (a5)--(dest5);
    \draw[move] (a4)--(dest4);
   \end{tikzpicture}
   \subcaption{Priority inheritance}
   \label{fig:bt:init}
  \end{minipage}
  &
  \begin{minipage}[t]{0.45\hsize}
   \centering
   \begin{tikzpicture}
    \node[vertex](v1) at (0, 0) {};
    \node[vertex,right=0cm of v1](v2) {};
    \node[vertex,right=0cm of v2](v3) {};
    \node[vertex,right=0cm of v3](v4) {};
    \node[vertex,above=0cm of v1](v5) {};
    \node[vertex,right=0cm of v5](v6) {};
    \node[vertex,right=0cm of v6](v7) {};
    \node[vertex,right=0cm of v7](v8) {};
    \node[agenthigh](a7) at (v6) {$a_7$};
    \node[agent](a6) at (v7) {$a_6$};
    \node[agentchange](a5) at (v3) {$a_5$};
    \node[agent](a4) at (v4) {$a_4$};
    \node[agentlow](a3) at (v8) {$a_3$};
    \node[agent](a1) at (v2) {$a_1$};
    \node[below=0.15cm of a1]() {\scriptsize low (as \textbf{high})};
    \node[agent](a2) at (v1) {$a_2$};
    \draw[bt] (a3)--(a4);
    \draw[bt] (a4)--(a5);
    \node[dest] (dest5) at (a1) {};
    \node[dest] (dest1) at (a2) {};
    \node[smalldest] (dest2) at (v5) {};
    \draw[move] (a5)--(dest5);
    \draw[move] (a1)--(dest1);
    \draw[move] (a2)--(dest2);
   \end{tikzpicture}
   \subcaption{Backtracking and priority inheritance again}
   \label{fig:bt:btpi}
  \end{minipage}
  \\
  \\
  \begin{minipage}[t]{0.45\hsize}
   \centering
   \begin{tikzpicture}
    \node[vertex](v1) at (0, 0) {};
    \node[vertex,right=0cm of v1](v2) {};
    \node[vertex,right=0cm of v2](v3) {};
    \node[vertex,right=0cm of v3](v4) {};
    \node[vertex,above=0cm of v1](v5) {};
    \node[vertex,right=0cm of v5](v6) {};
    \node[vertex,right=0cm of v6](v7) {};
    \node[vertex,right=0cm of v7](v8) {};
    \node[agenthigh](a7) at (v6) {$a_7$};
    \node[agent](a6) at (v7) {$a_6$};
    \node[agentchange](a5) at (v3) {$a_5$};
    \node[agent](a4) at (v4) {$a_4$};
    \node[agent](a3) at (v8) {$a_3$};
    \node[agent](a1) at (v2) {$a_1$};
    \node[agent](a2) at (v1) {$a_2$};
    \draw[bt] (a2)--(a1);
    \draw[bt] (a1)--(a5);
    \draw[bt] (a5)--(a6);
    \draw[bt] (a6)--(a7);
   \end{tikzpicture}
   \subcaption{Backtracking}
   \label{fig:bt:bt}
  \end{minipage}
  &
  \begin{minipage}[t]{0.45\hsize}
   \centering
   \begin{tikzpicture}
    \node[vertex](v1) at (0, 0) {};
    \node[vertex,right=0cm of v1](v2) {};
    \node[vertex,right=0cm of v2](v3) {};
    \node[vertex,right=0cm of v3](v4) {};
    \node[vertex,above=0cm of v1](v5) {};
    \node[vertex,right=0cm of v5](v6) {};
    \node[vertex,right=0cm of v6](v7) {};
    \node[vertex,right=0cm of v7](v8) {};
    \node[agenthigh](a7) at (v7) {$a_7$};
    \node[agent](a6) at (v3) {$a_6$};
    \node[agent](a5) at (v2) {$a_5$};
    \node[agent](a4) at (v4) {$a_4$};
    \node[agentlow](a3) at (v8) {$a_3$};
    \node[agent](a1) at (v1) {$a_1$};
    \node[agent](a2) at (v5) {$a_2$};
   \end{tikzpicture}
   \subcaption{one timestep later}
   \label{fig:bt:result}
  \end{minipage}
 \end{tabular}
 \caption{
   \textbf{Example of PIBT.}
   The flow of back tracking is indicated by double-lined arrows.
   Because $a_3$ is stuck (\ref{fig:bt:init}), backtracking returns as invalid to $a_4$, and subsequently to $a_5$.
   $a_5$ executes other priority inheritance to $a_1$ (\ref{fig:bt:btpi}).
   $a_1, a_5, a_6$ and $a_7$ wait for the results of back tracking (\ref{fig:bt:bt}) and then start moving (\ref{fig:bt:result}).
 }
 \label{fig:bt}
\end{figure}

\subsubsection{Backtracking}
Priority inheritance deals effectively with priority inversion; however, it does not completely ensure deadlock freedom. For instance, as shown in Fig.~\ref{fig:bt:init}, an agent $a_3$ cannot escape as a result of consecutive priority inheritances ($a_7 \rightarrow a_6 \rightarrow a_5 \rightarrow a_4$).

The solution relies on \emph{backtracking}; any agent~$a_i$ that executes priority inheritance must wait for an outcome, valid or invalid. If valid, $a_i$ successfully moves to the desired node. Otherwise, it must request a different node, excluding:
(1)~nodes requested by a higher priority agent, and
(2)~nodes having already returned an invalid outcome.
Upon finding no valid or unoccupied nodes, $a_i$ sends back an invalid outcome to the agent from which it inherited its priority.

For instance, in Fig.~\ref{fig:bt:btpi}, $a_3$ sends ``invalid'' back to $a_4$ as the outcome of the request received previously from $a_4$ (Fig.~\ref{fig:bt:init}). $a_4$ tries to replan its next location; however, in the absence of other alternatives, in turn sends invalid to $a_5$. Because $a_1$ has lower priority, $a_5$ can let $a_1$ inherit its priority as an alternative, which leads to a valid outcome (Fig.~\ref{fig:bt:bt}), and agents can move (Fig.~\ref{fig:bt:result}).

By combining priority inheritance and backtracking, it is ensured that all agents are assigned to the next locations without collisions.

\subsection{Algorithm}
We formalize the concept in Algorithm~\ref{algo:pibt}, which consists of (1)~a top-level procedure [Lines~\ref{algo:pibt:assign-priority}--\ref{algo:pibt:loop-end}] calling (2) a recursive procedure [Lines~\ref{algo:pibt:func-pibt-start}--\ref{algo:pibt:func-pibt-end}] that executes priority inheritance and backtracking.

\begin{algorithm}[t!]
  \caption{PIBT}
  \label{algo:pibt}
  \begin{algorithmic}[1]
  \item[\textbf{Input}:~graph $G$, agents $A$, starts $\{ s_1, \ldots s_n \}$, goals $\{ g_1, \ldots g_n \}$]
  \item[\textbf{Output}:~paths $\{ \path{1}, \ldots, \path{n} \}$]
  \item[\textbf{Preface}:~\loc{i}{t} is assumed to be initialized with $\bot$]
    \State $\loc{i}{0} \leftarrow s_i$: for each agent $a_i \in A$
    \State $p_i \leftarrow \epsilon_i$: for each agent $a_i \in A$ s.t. $\epsilon_i \in [0, 1)$ and $\epsilon_i \neq \epsilon_j (i\neq j)$
    \label{algo:pibt:init-priorities}
    \Comment setup priorities
    \bigskip
  \item[(for each timestep $t=1, 2, \ldots$ until terminates, repeat the following)]
    \State $p_i \leftarrow$ \textbf{if} $\loc{i}{t} = g_i$ \textbf{then} $\epsilon_i$ \textbf{else} $p_i + 1$: for each agent $a_i \in A$
    \label{algo:pibt:assign-priority}
    \Comment update priorities
    \State sort $A$ in decreasing order of priorities $p_i$
    \label{algo:pibt:sort-agents}
    \For{$a_i \in A$}
    \label{algo:pibt:loop-start}
    \IFSINGLE{$\loc{i}{t+1} = \bot$}{$\funcname{PIBT}(a_i, \bot)$}
    \label{algo:pibt:call-func}
    \EndFor
    \label{algo:pibt:loop-end}
    \bigskip
    \Procedure{$\funcname{PIBT}$}{$a_i$, $a_j$}
    \Comment we use $t$ implicitly
    \label{algo:pibt:func-pibt-start}
    \State $C \leftarrow \neigh{\loc{i}{t}} \cup \{ \loc{i}{t} \}$
    \label{algo:pibt:candidates}
    \State sort $C$ in increasing order of $\dist{u}{g_i}$ where $u \in C$
    \label{algo:pibt:sort-candidates}
    \Comment tie-break: presence of agents
    \For{$v \in C$}
    \label{algo:pibt:func-loop-start}
    \IFSINGLE{$\exists a_k \in A~\text{s.t.}~\loc{k}{t+1}=v$}{\textbf{continue}}
    \label{algo:pibt:avoid-vertex-conflict}
    \Comment avoid vertex conflict
    \IFSINGLE{$a_j \neq \bot \land \loc{j}{t}=v$}{\textbf{continue}}
    \label{algo:pibt:avoid-swap-conflict}
    \Comment avoid swap conflict
    \State $\loc{i}{t+1} \leftarrow v$
    \label{algo:pibt:reserve}
    \Comment reserve
    \If{$\exists a_k \in A~\text{s.t.}~\loc{k}{t}=v \land \loc{k}{t+1}=\bot$}
    \label{algo:pibt:check-another-agent-start}
    \IFSINGLE{$\funcname{PIBT}(a_k, a_i)$ is \invalid}{\textbf{continue}}
    \label{algo:pibt:priority-inheritance}
    \Comment replanning
    \EndIf
    \label{algo:pibt:check-another-agent-end}
    \State \Return \valid
    \label{algo:pibt:valid}
    \EndFor
    \label{algo:pibt:func-loop-end}

    \State $\loc{i}{t+1} \leftarrow \loc{i}{t}$
    \label{algo:pibt:stay}
    \State \Return \invalid
    \label{algo:pibt:invalid}
    \EndProcedure
    \label{algo:pibt:func-pibt-end}
 \end{algorithmic}
\end{algorithm}

\paragraph{Top-level Procedure}
PIBT first updates the priority $p_i \in \mathbb{R}_{\geq 0}$ for each agent $a_i$ [Line~\ref{algo:pibt:assign-priority}]. The rule is simple; increment $p_i$ when $a_i$ has not reached its goal $g_i$; otherwise, reset $p_i$ to $\epsilon_i \in [0, 1)$. The tie-breaker $\epsilon_i$, which is distinct for agents, has the role of keeping $p_i$ unique among agents. Subsequently, according to priorities, the algorithm assigns locations to each agent $a_i$ without a next location \loc{i}{t+1} via the procedure \funcname{PIBT} [Line~\ref{algo:pibt:call-func}]. The termination of the algorithm is explained later.

\paragraph{Recursive Procedure}
The procedure \funcname{PIBT} implements the concept of priority inheritance and backtracking. It has two arguments: the agent $a_i$ making a decision and an agent $a_j$ from which $a_i$ inherits its priority, or $\bot$ if there is none. The procedure returns \invalid if $a_i$ becomes a stuck agent like the red agent $a_3$ in Fig.~\ref{fig:bt}, otherwise returns \valid. Once this procedure is called, $a_i$ must determine \loc{i}{t+1} from the ordered set of candidate nodes $C$. Here, $C$ consists of \loc{i}{t} and its neighbors [Line~\ref{algo:pibt:candidates}], sorted in increasing order of distance from the goal~$g_i$ [Line~\ref{algo:pibt:sort-candidates}]. To avoid unnecessary priority inheritance, the presence (or absence) of an agent is used as a tie-breaking rule. $C$ is filtered, excluding two kinds of nodes: (1)~nodes already requested by someone (to avoid vertex conflict) [Line~\ref{algo:pibt:avoid-vertex-conflict}] and (2)~\loc{j}{t} if $a_j \neq \bot$ (to avoid swap conflict) [Line~\ref{algo:pibt:avoid-swap-conflict}]. If no nodes remain in $C$, this is the case of a stuck agent; then, $a_i$ must stay at the current location and the procedure returns \invalid [Lines~\ref{algo:pibt:stay}--\ref{algo:pibt:invalid}].

For each node $v \in C$, the procedure checks whether $a_i$ can move to $v$ in the next timestep. After reserving $v$ for $a_i$ [Line~\ref{algo:pibt:reserve}], the priority inheritance occurs from $a_i$ to $a_k$ when another agent $a_k$, which has not yet determined the next location, occupies $v$ [Lines~\ref{algo:pibt:check-another-agent-start}--\ref{algo:pibt:check-another-agent-end}]. If $a_k$ returns \invalid, the next location for $a_i$ is replanned [Line~\ref{algo:pibt:priority-inheritance}], or else, $a_i$ finishes the planning and returns \valid, as well as that $v$ is unoccupied by others [Line~\ref{algo:pibt:valid}].

A step-by-step example of Algorithm~\ref{algo:pibt} is illustrated in Fig.~\ref{fig:running-example}.
\input{fig/running-example}

\subsection{Theoretical Analysis}

\subsubsection{Reachability}
Here, it is proven that all agents eventually reach their respective destination in graphs with adequate properties (e.g., biconnected graphs). The proof relies on the following Lemma~\ref{thrm:local-move}, which states that the agent with the highest priority at each timestep is always assigned to its desired node, assuming that it can always move along a simple cycle.

\begin{figure}[t]
  \centering
  \tikzset{
    invalid/.style={bt,color={rgb:red,3;green,0;blue,0}},
  }
  \begin{tikzpicture}
    \node[vertex](v1) at (0.0, 0.0) {};
    \node[vertex](v2) at (1.8, 0.4) {};
    \node[vertex](v3) at (3.6, 0.2) {};
    \node[vertex](v4) at (0.4, -1.6) {};
    \draw[line](v1)--(v2);
    \draw[line](v2)--(v3);
    \draw[line](v4)--(v1);
    \draw[line, dashed](v3) to[out=330,in=290] (v4);
    \node[agenthigh](a1) at (v1) {$a_1$};
    \node[agent](a2) at (v2) {$a_2$};
    \node[agent](a3) at (v3) {$a_3$};
    \node[agentgreen](ak) at (v4) {$a_k$};
    %
    \draw[pi](0.3, 0.3)--(1.4, 0.6);
    \draw[pi](1.8, 0.8)--(1.7, 1.2);
    \draw[pi](1.7, 0.0)--(1.8, -0.5);
    \draw[pi](2.2, 0.5)--(3.2, 0.4);
    \draw[pi](3.9, 0.2)--(4.6, -0.2);
    \draw[pi](4.5, -0.9)--(3.8, -1.7);
    \draw[pi](3.2, -2.2)--(2.2, -2.6);
    \draw[pi](0.8, -2.8)--(0.1, -2.0);
    %
    \draw[bt](0.7, -2.0)--(1.3, -2.3);
    \draw[bt](2.0, -2.2)--(2.7, -1.9);
    \draw[bt](3.4, -1.6)--(3.8, -1.1);
    \draw[bt](4.3, -0.5)--(3.8, -0.1);
    \draw[bt](3.2, 0)--(2.2, 0.1);
    \draw[bt](1.4, 0)--(0.3, -0.3);
    %
    \draw[invalid](2.2, -0.5)--(2.0, 0.0);
    \draw[invalid](2.0, 1.3)--(2.1, 0.8);
    \draw[invalid](3.2, -0.8)--(3.4, -0.2);
    \draw[invalid](3.8, 1.1)--(3.7, 0.6);
    \node[left=0cm of v1](){\loc{1}{t}};
    \node[above left=-0.2cm of v2](){$v^\ast$};
    {
      \draw[pi](0, -1.2)--(-0.2, -0.4);
      \node[anchor=east]() at (-0.1, -0.8) {\scriptsize this action is sure to succeed};
    }
    \node[anchor=west]() at (4.6, -0.6) {$\bm{C}${\scriptsize (simple cycle)}};
    \node[anchor=west]() at (4.2, -1.3) {\scriptsize priority inheritance};
    {
      \node[anchor=west]() at (1.5, -1.5) {\scriptsize backtracking};
      \node[anchor=west]() at (1.5, -1.75) {\scriptsize (valid)};
    }
    {
      \node[anchor=west]() at (3.9, 1.10) {\scriptsize backtracking};
      \node[anchor=west]() at (3.9, 0.85) {\scriptsize (invalid)};
    }
    {
      \node[anchor=east]() at (0.2, 0.7) {\scriptsize \textbf{highest}};
    }
  \end{tikzpicture}
  \vspace{-1cm}
  \caption{
    \textbf{Proof sketch of Lemma~\ref{thrm:local-move}.}
    The agent with the highest priority ($a_1$) can relocate to $v^\ast$ if there exists a simple cycle $\bm{C} = \left(\loc{1}{t}, v^\ast, \ldots\right)$.
  }
 \label{fig:proof-sketch}
\end{figure}

\begin{lemma}
Let $a_1$ denote the agent with highest priority at timestep~$t$ and $v^\ast$ the nearest node to $g_1$ among neighbors of \loc{1}{t}. If a simple cycle $\bm{C}=\left( \loc{1}{t}, v^\ast, \ldots \right)$ exists, Algorithm~\ref{algo:pibt} assigns $v^\ast$ to $a_1$ as \loc{1}{t+1}.
  \label{thrm:local-move}
\end{lemma}
\begin{proof}
Fig.~\ref{fig:proof-sketch} visualizes the following.

Assume that the top-level procedure calls $\funcname{PIBT}(a_1, \bot)$ at Line~\ref{algo:pibt:call-func} for the agent with highest priority. At this phase, \loc{i}{t+1} is $\bot$ for any agent $a_i$, i.e., the next location has not been assigned yet. Consequently, $a_1$ selects $v^\ast$ as a target node $v$ in the first iteration of Lines~\ref{algo:pibt:loop-start}--\ref{algo:pibt:loop-end}. Then, it remains to prove that Line~\ref{algo:pibt:priority-inheritance}, priority inheritance, never returns \invalid when another agent $a_2$ occupies $v^\ast$. In the other cases, $a_1$ evident gets $v^\ast$.

Next, assume that $\funcname{PIBT}(a_2, a_1)$ is called from the original $\funcname{PIBT}(a_1, \bot)$. The existence of the cycle $\bm{C} = (\loc{1}{t}, v^\ast)$ guarantees that $a_2$ has a non-empty set of candidate nodes $C_2$ for the next timestep without colliding with $a_1$. During the iterations of Lines~\ref{algo:pibt:func-loop-start}--\ref{algo:pibt:func-loop-end}, $\funcname{PIBT}(a_2, a_1)$ returns \valid when an unoccupied node $v \in C_2$ at timestep $t$ is selected. This forms the basis of the remaining proof by induction.

Following this, suppose that $\funcname{PIBT}(a_{i}, a_{i-1})$ is called before $\funcname{PIBT}(a_2, a_1)$ returns any value. Similar to the case of $a_2$, $\funcname{PIBT}(a_{i}, a_{i-1})$ must return \valid when one of the surrounding nodes of \loc{i}{t} is unoccupied at timestep $t$. In addition, as distinct from $a_2$, which must avoid swap conflict with $a_1$, $a_i$ can select $\loc{1}{t}$ and return \valid when \loc{1}{t} is neighbor to $\loc{i}{t}$. As a result, $\funcname{PIBT}(a_{i-1}, a_{i-2})$, which calls $\funcname{PIBT}(a_{i}, a_{i-1})$ also returns \valid and eventually, $\funcname{PIBT}(a_2, a_1)$ returns \valid.

Next, by contradiction, suppose that $\funcname{PIBT}(a_2, a_1)$ returns \invalid. This assumption means that, for all agents $a_j (\neq a_1)$ neighboring $a_2$ at timestep $t$, $\funcname{PIBT}(a_j, \cdot)$ returns \invalid. This is the same for $a_j$, and so forth. Meanwhile, the existence of $\bm{C}$ indicates that at least one agent $a_k \neq a_2$ such that $\loc{k}{t} \in \bm{C}$ had initially at least one free neighbor node. Even though all nodes in $\bm{C}$ are occupied, the agent on the last node of $\bm{C}$ has a free neighbor node, i.e., \loc{1}{t}. This is a contradiction; $\funcname{PIBT}(a_k, \cdot)$ returns \valid; consequently, $\funcname{PIBT}(a_2, a_1)$ returns \valid.
\end{proof}

\begin{theorem}
  Given an MAPF instance such that $G$ has a simple cycle for all pairs of adjacent nodes, PIBT (Algorithm~\ref{algo:pibt}) generates a set of conflict-free paths $\{ \path{1}, \ldots, \path{n} \}$ such that for any agent $a_i \in A$, $\loc{i}{0} = s_i$ and there is a timestep $t \leq \mathit{diam}(G)\cdot |A|$ such that $\loc{i}{t} = g_i$.
  \label{thrm:reachability}
\end{theorem}
\begin{proof}
  From Lemma~\ref{thrm:local-move}, the agent with highest priority reaches its own goal within $\mathit{diam}(G)$ timesteps. Based on the update rule of the priority $p_i$ for an agent $a_i$, once some agent $a_i$ has reached its goal, $p_i$ is reset and is lower than the priority of all other agents who have not reached their goal yet. These agents see their priority increase by one. As long as such agents remain, exactly one of them must have the highest priority. In turn, this agent reaches its own goal after at most $\mathit{diam}(G)$ timesteps. This repeats until all agents have reached their goal at least once, which takes at most $\mathit{diam}(G)\cdot |A|$ timesteps in total.
\end{proof}

Hereafter, this property is referred to as \emph{reachability}, that is, all agents eventually reach their goal. Reachability differs from completeness for one-shot MAPF; it is never ensured that all agents reach their goal \emph{simultaneously}; hence, PIBT is not ensured to solve conventional MAPF, defined in Section~\ref{sec:def:mapf}. Alternatively, with the graph condition of Theorem~\ref{thrm:reachability}, PIBT is complete for the MAPF variant where agents need not necessarily stay at their goals.

A typical example that satisfies the aforementioned graph condition is the biconnected undirected graph. The opposite is not true, however, and Theorem~\ref{thrm:reachability} is expressed more generally to consider directed graphs. For instance, a directed ring satisfies the condition even though it is not biconnected.

Without the graph condition in Theorem~\ref{thrm:reachability}, several agents may remain in the same vertices permanently and may not reach their goals.
Figure~\ref{fig:failure} shows such an example.
In practice, PIBT needs a pre-defined maximum planning timestep to detect such failure cases.
When PIBT reaches this timestep, it stops planning and reports a planning failure.
See also Section~\ref{sec:pibt-mapf}.

\begin{figure}[t]
  \newcommand{\colwidth}{0.3\hsize}
  \centering
  \begin{tabular}{cc}
    \begin{minipage}[t]{\colwidth}
      \centering
      \begin{tikzpicture}
        \node[vertex](v1) at (0, 0) {};
        \node[vertex,right=0cm of v1](v2) {};
        \node[vertex,right=0cm of v2](v3) {};
        \node[vertex,above=0cm of v2](v4) {};
        \node[agenthigh](a1) at (v2) {$a_1$};
        \node[agentlow](a2) at (v3) {$a_2$};
        \node[dest] (dest1) at (v3) {};
        \draw[move](a1)--(dest1);
      \end{tikzpicture}
      \subcaption{Priority inheritance}
      \label{fig:failure:1}
    \end{minipage}
    &
      \begin{minipage}[t]{\colwidth}
      \centering
      \begin{tikzpicture}
        \node[vertex](v1) at (0, 0) {};
        \node[vertex,right=0cm of v1](v2) {};
        \node[vertex,right=0cm of v2](v3) {};
        \node[vertex,above=0cm of v2](v4) {};
        \node[agenthigh](a1) at (v2) {$a_1$};
        \node[agentlow](a2) at (v3) {$a_2$};
        \draw[bt](a2)--(a1);
      \end{tikzpicture}
      \subcaption{Backtracking}
      \label{fig:failure:2}
    \end{minipage}
  \end{tabular}
  \caption{
    \textbf{Failure case of PIBT on a graph without cycles.}
    Assume that an agent $a_1$'s goal is the location at another agent $a_2$.
    The priority inheritance to $a_2$ (\ref{fig:failure:1}) is ``invalid'' (\ref{fig:failure:2}) because $a_2$ has no escape node.
    Then, $a_1$ does replanning, however, it selects the current location as the next location since it is the next nearest vertex towards the goal.
    As a result, this situation will be held beyond the current timestep and there will be no progress.
  }
  \label{fig:failure}
\end{figure}

\subsubsection{Complexity Analysis}
Now we consider the time complexity of PIBT. Let $F$ be the maximum time required for sorting candidate nodes [Line~\ref{algo:pibt:sort-candidates}]. $F$ depends on both $G$ and the distance computation for the goals.

\begin{proposition}
  \label{thrm:time-complexity}
  The time complexity of PIBT in one timestep is $O\left(|A|\cdot(\Delta(G) + F + \lg{|A|})\right)$.
\end{proposition}
\begin{proof}
  Within one timestep, the procedure \funcname{PIBT} is called exactly $|A|$ times, once for each agent. This is because $\funcname{PIBT}(a_i, \cdot)$ is called if and only if an agent $a_i$ has not yet determined the next location (that is, when $\loc{i}{t+1} = \bot$) [Lines~\ref{algo:pibt:call-func}, \ref{algo:pibt:priority-inheritance}] but $\funcname{PIBT}(a_i, \cdot)$ assigns \loc{i}{t+1} [Line~\ref{algo:pibt:reserve}] before calling another $\funcname{PIBT}(a_k, a_i)$ for priority inheritance. The loop of Line~\ref{algo:pibt:func-loop-start}--\ref{algo:pibt:func-loop-end} iterates at most $\Delta(G) + 1$ times. Each operation in the loop is performed in constant time. From the assumption, Line~\ref{algo:pibt:sort-candidates} requires $O(F)$. As a result, the procedure \funcname{PIBT} requires $O(|A|\cdot(\Delta(G) + F))$.

  Line~\ref{algo:pibt:assign-priority} requires $O(|A|)$. Line~\ref{algo:pibt:sort-agents} requires $O(|A|\lg{|A|})$. In total, PIBT in one timestep requires $O\left(|A|\cdot(\Delta(G) + F + \lg{|A|})\right)$.
\end{proof}

The analysis is further continued for specific conditions. Assume that a distance table for the goal of each agent is computed by breadth-first search prior to executing PIBT, where the overhead is $O(|A|\cdot |E|)$. Then, $F$ will be solely sorting candidate nodes following the table, resulting in $O(\Delta(G)\lg{\Delta(G)})$. Consequently, PIBT in one timestep is $O\left(|A|\cdot(\Delta(G)\lg{\Delta(G)} + \lg{|A|})\right)$. Assume further that $G$ is a 4-connected grid, as commonly used in MAPF studies. Here, instead of $\Delta(G)$, consider $\Delta(G)+1$ for accurate analysis. Then, $(\Delta(G)+1)\lg{(\Delta(G)+1)} \simeq 11.6$. As a result, without a huge team of agents ($|A| < 3125$), the first term $|A|\cdot\Delta(G)\lg{\Delta(G)}$ is dominant; otherwise, the second term $|A|\lg{|A|}$ is dominant. In either case, \emph{PIBT can be said to be computationally very inexpensive}.

The small time complexity provides several advantages. For instance, PIBT can address large instances for both $A$ and $G$, which other MAPF algorithms cannot solve within a realistic timeframe. Another advantage is an application to \emph{anytime planning}, a scheme that yields a feasible solution whenever interrupted but gradually improves solution quality as time goes by. Anytime planning is attractive, particularly in on-demand situations where the deliberation time is finite. A previous study realizes anytime planning for MAPF by refining known solutions iteratively~\cite{okumura2021iterative}. For such an approach, obtaining initial solutions as quickly as possible is key because we can leave much time for refinement. In this sense, PIBT is a good choice to obtain an initial solution. In contrast, the anytime approach cannot be realized by ``slow'' solvers because they may not provide solutions within a time limit. Lastly, given a certain timestep, the runtime of PIBT is \emph{predictable}, which is a fundamental characteristic in real-time planning. The power of PIBT is reflected in the experiment.

\subsection{Application to Specific Problems}
We now present how to adapt PIBT to solve typical pathfinding problems of multiple agents, namely, MAPF and MAPD.

\subsubsection{Application to MAPF}
\label{sec:pibt-mapf}
For convenience, \emph{configuration} refers to a set of locations of all agents at a given timestep. For instance, the initial configuration is $\left(s_1, \ldots, s_n\right)$.

\paragraph{Termination}
So far, when the PIBT run has been completed, has not been specified. In the conventional MAPF, PIBT is assumed to run until it reaches the goal configuration $(g_1, \ldots, g_n)$; otherwise, PIBT ``fails'' to solve the MAPF instance when it reaches the pre-defined timestep.

\paragraph{Distance evaluation}
In PIBT, for each timestep, each agent has to evaluate distances from the surrounding nodes to its goal [Line~\ref{algo:pibt:sort-candidates} of Algorithm~\ref{algo:pibt}].
This operation could be implemented by calling \astar on demand but it could also be a bottleneck.
Instead, PIBT can save computation time by preparing distance tables from each agent's goal.
This is computed by breadth-first search from the goals with an overhead of $O(|A|\cdot |E|)$.
Indeed, this is the main reason for the speedup of the implementation from our previous version of PIBT~\cite{okumura2019priority}.

\paragraph{Extension for one-shot MAPF}
Even if $G$ satisfies the condition for reachability, PIBT does not ensure that all agents reach their goal simultaneously, which is a requirement of conventional MAPF. In fact, with naive PIBT, a certain kind of livelock situation was confirmed in our experiment.
This motivates the development of PIBT$^+$, a framework that enhances known MAPF solvers, presented in Algorithm~\ref{algo:pibt-plus}.
The main idea behind PIBT$^+$ is to make problem instances easier for the MAPF solver by bringing agents near their destinations using PIBT.
In this sense, we say that the target MAPF solver is a \emph{complement} solver.

\begin{algorithm}[ht!]
  \caption{PIBT$^+$}
  \label{algo:pibt-plus}
  \begin{algorithmic}[1]
  \item[\textbf{Input}:~graph $G$, agents $A$, starts $\{ s_1, \ldots s_n \}$, goals $\{ g_1, \ldots g_n \}$]
  \item[\textbf{Output}:~paths $\paths = \{ \path{1}, \ldots, \path{n} \}$]
    \State $\tmin \leftarrow \max_{a_j \in A} \dist{\loc{j}{0}}{g_j}$
    \State Set initial priorities $p_i$ to agents in descending order of distance from their initial location to their goals by adjusting the tie-breaker $\epsilon_i$ at Line~\ref{algo:pibt:init-priorities} in Alg.~\ref{algo:pibt}
    \State $C_t \leftarrow$~configuration at \tmin in \paths
    \label{algo:pibt-plus:run-pibt}
    \State $\paths \leftarrow$ PIBT (Alg.~\ref{algo:pibt}) until timestep \tmin
    \IFSINGLE{$C_t = (g_1, \ldots, g_n)$}{\Return \paths}
    \State $\paths' \leftarrow$ other MAPF algorithm (\emph{complement solver}), taking $C_t$ as the initial configuration
    \label{algo:pibt-plus:complement-solver}
    \State \Return a concatenation of $\paths$ and $\paths'$
 \end{algorithmic}
\end{algorithm}

The concept of Algorithm~\ref{algo:pibt-plus} is as follows. \tmin is the minimum number of timesteps needed for solutions from the initial configuration to the goal configuration. Because PIBT can be regarded as prioritized planning, the agent with the longest initial distance reaches its goal following the shortest path by providing priorities according to the initial distance. This implies that there is a chance that all agents reach their goals at \tmin. The remaining problem is relatively easy compared to the original because most agents are expected to have already reached their goals.
\begin{theorem}
  PIBT$^+$ is complete for MAPF on undirected graphs as long as the completeness conditions of the complement solver are satisfied.
\end{theorem}
\begin{proof}
  Let $C_s, C_t, C_g$ be the initial configuration, the configuration at \tmin planned by Line~\ref{algo:pibt-plus:run-pibt} in Algorithm~\ref{algo:pibt-plus}, and the goal configuration, respectively. There exists a solution from $C_t$ to $C_s$ because PIBT computes the plan from $C_s$ to $C_t$. If the original problem is solvable, there exists a solution from $C_s$ to $C_g$, meaning that, at least one solution exists from $C_t$ to $C_g$. According to this assumption, PIBT$^+$ finally uses the complete solver from $C_t$ to $C_g$, and the complete solver must return a solution. The above satisfies the statement.
\end{proof}

PIBT$^+$ is essentially helpful for situations where it is difficult to obtain solutions through a direct adaptation of the complement solver but where plausible solutions are sought in a short time.
As a complement solver, it is desirable to use complete algorithms such as rule-based, search-based, or compiling-based approaches. We note that PIBT$^+$ does not guarantee a better solution nor finish planning faster than using the complement solver from the beginning. On the other hand, since the transition from the initial configuration to the configuration at \tmin is reversible, the additional burdens of sum-of-cost and makespan from optimal ones can be, respectively, at most $2|A|\cdot\tmin$ and $2\tmin$. Similarly, the additional burden of time complexity can be $O(\tmin\cdot |A|\cdot(\Delta(G) + F + \log{|A|})$ from the consequence of Proposition~\ref{thrm:time-complexity}.

\subsubsection{Applying to MAPD}
The MAPD problem is a typical example of online and lifelong scenarios requiring task allocation. We here describe how PIBT adapts to MAPD.

To design a complete MAPD algorithm, we must guarantee two claims: (1)~all non-assigned tasks are eventually assigned, and (2)~all assigned tasks are eventually completed, i.e., a task-assigned agent must visit pickup and delivery locations in order. For simplification, consider assigning a task to an agent only when the agent is at the pickup location. The remaining work for the agent is to visit the delivery location.

PIBT (Algorithm~\ref{algo:pibt}) is convenient for the second claim due to the following reason. Recall that Lemma~\ref{thrm:local-move} states that, for each timestep, the agent with the highest priority can always move to an arbitrary neighbor node. Using this lemma, ensuring that a task-assigned agent completes the task in finite time is straightforward. This is achieved with a special prioritization scheme that satisfies the two conditions: (1)~all task-assigned agents have higher priorities than free agents, and (2)~every task-assigned agent eventually gets the highest priority and holds it until the task is completed.

The first claim, all non-assigned tasks are eventually assigned, is achieved by various approaches. We here take a simple approach, i.e., for each timestep, all free agents set their goals to the nearest pickup location of non-assigned tasks. Assume that every free agent eventually gets the highest priority among free agents and holds the highest until it reaches a goal (a pickup location). Then, this approach satisfies the first claim because all task-assigned agents eventually become free due to the second claim.

\begin{algorithm}[ht!]
  \caption{PIBT for MAPD}
  \label{algo:pibt-mapd}
  \begin{algorithmic}[1]
  \item[\textbf{Input}:~graph $G$, agents $A$, initial locations $\{\loc{1}{0}, \ldots, \loc{n}{0}\}$, task stream $\Gamma$]
  \item[\textbf{Output}:~task assignment and paths $\{ \path{1}, \ldots, \path{n} \}$]
    \item[]{\scriptsize\textit{(repeat below for each timestep $t = 0, 1, \ldots$ until all tasks are completed)}}
    \For{$a_i \in A$}
    \label{algo:pibt-mapd:task-assign}
    \IFSINGLE{$a_i$ is assigned to a task $\tau = (u, v)$}{$g_i \leftarrow v$; \textbf{continue}}
    \State Let $\tau' = (u', v')$ be an unassigned task from $\Gamma$ that minimizes $\dist{\loc{i}{t}}{u'}$
    \If{no such $\tau'$ exists}
    \State $g_i \leftarrow \loc{i}{t}$
    \ElsIf{$u' = \loc{i}{t}$}
    \State assign $\tau'$ to $a_i$; remove $\tau$ from $\Gamma$; $g_i \leftarrow v'$
    \Else
    \State $g_i \leftarrow u'$
    \EndIf
    \EndFor
    \label{algo:pibt-mapd:task-assign-end}
    \State $\mathcal{S} \leftarrow \{ \loc{1}{t}, \ldots, \loc{n}{t} \}$; $\mathcal{G} \leftarrow \{ g_1, \ldots, g_n \}$
    \label{algo:pibt-mapd:start-goal}
    \State $\loc{1}{t+1}, \ldots, \loc{n}{t+1} \leftarrow$ Algorithm~\ref{algo:pibt} for one-timestep with starts $\mathcal{S}$ and goals $\mathcal{G}$, while adding a prioritization rule such that all task-assigned agents have higher ones than those of free agents
    \label{algo:pibt-mapd:pibt}
    \For{each agent $a_i \in A$ assigned to a task $\tau = (v, u)$}
    \IFSINGLE{$\loc{i}{t+1} = u$}{make $a_i$ free}
    \Comment $\tau$ is completed
    \EndFor
 \end{algorithmic}
\end{algorithm}

Algorithm~\ref{algo:pibt-mapd} realizes the aforementioned concept to solve MAPD. For each timestep, the algorithm first assigns tasks while updating the agents' goals [Lines~\ref{algo:pibt-mapd:task-assign}--\ref{algo:pibt-mapd:task-assign-end}]. Then, the algorithm uses PIBT to plan locations for the next timestep [Lines~\ref{algo:pibt-mapd:start-goal}--\ref{algo:pibt-mapd:pibt}]. As discussed earlier, for each timestep, Algorithm~\ref{algo:pibt-mapd} prioritizes the task-assigned agents than free agents. As a secondary prioritization, it uses the original PIBT prioritization scheme (Line~\ref{algo:pibt:assign-priority} of Algorithm~\ref{algo:pibt}) as it is.

The following theorem is a consequence of the aforementioned discussion and its reachability.
\begin{theorem}
  Algorithm~\ref{algo:pibt-mapd} is complete for an MAPD problem when $G$ has a simple cycle for all pairs of adjacent nodes.
\end{theorem}

\paragraph{Comparison of our Proposed Approach with an Existing Complete Approach}
We later compare Algorithm~\ref{algo:pibt-mapd} with the TP algorithm~\cite{ma2017lifelong} experimentally. Before that, we here provide a qualitative discussion as follows. TP is complete for MAPD in a limited situation. First, it requires locations that never be either pickup or delivery locations for any tasks, called non-task endpoints. The non-task endpoints are required at least for the number of agents. Then, for any two endpoints, a path must exist without traversing any other endpoints. Here, endpoints consist of non-task endpoints and candidates of pickup/delivery locations. Compared to TP, PIBT does not require such conditions and works in a wide range of situations as long as the graph condition is satisfied.

\paragraph{Other Task Allocation}
We took a simple and greedy task allocation process; however, more aggressive approaches are applicable, e.g., incorporating linear cost-optimal assignment~\cite{kuhn1955hungarian} or bottleneck assignment~\cite{gross1959bottleneck}. Doing so can improve solution qualities but potentially requires more computation time.

\subsection{Decentralized Online Planning}
\label{sec:decentralization}
Here, the decentralization of PIBT implementations is briefly discussed. In particular, we focus on the \emph{online planning} of PIBT, that is, agents repeat a set of single-timestep planning and move actions. Because PIBT is based on prioritized planning, adopting a decentralized context is realistic. The part of priority inheritance and backtracking is performed by the propagation of information. Further, PIBT relies on local interactions between agents, implying that agents are not required to know other agents' information far away from their current location.
To illustrate this, the concept of \emph{interacting agents}, a set of agents that must negotiate their planning, is introduced.

\paragraph{Interacting Agents}
When two agents are located within two hops of each other’s move, they may collide in the next timestep; then, they are said to be \emph{directly interacting} in that timestep. For an agent $a_i$, a group of \emph{interacting agents} $\mathcal{A}_i(t) \subseteq A$ is then defined by transitivity over direct interactions in timestep~$t$. Note that, given $a_i$ and $\mathcal{A}_i(t)$, for any other agent $a_j \in \mathcal{A}_i(t)$, we have $\mathcal{A}_j(t) = \mathcal{A}_i(t)$.
Whenever the context is obvious, $\mathcal{A}$ is directly used.

Because agents belonging to different groups cannot affect each other, path planning, the negotiation process using priority inheritance and backtracking can effectively occur in parallel. As a result, theoretically speaking, PIBT can be decentralized, relying only on local interactions, that is, it is sufficient that two agents in close proximity talk directly and utilize multi-hop communication; however, note that the groups of interacting agents are fully identified prior to starting a PIBT process for one timestep.

\paragraph{Communication}
Assume that interacting agents are fully detected at each timestep. Then, we analyze communication complexity, i.e., how many messages are required in PIBT. In a decentralized context, PIBT requires agents to know others' priorities before deciding on the next nodes. Usually, this requires $|\mathcal{A}|^2$ communication between agents; however, the update rule of the priority $p_i$ relaxes this effort, for instance, storing other agents' priorities and communicating only when the priority is reset. Therefore, the communication cost of PIBT mainly depends on the information propagation phase.

Next, consider this information propagation phase. In PIBT, communication between agents corresponds to calling \funcname{PIBT} (priority inheritance), and when the procedure returns a value (corresponding to backtracking). $\funcname{PIBT}(a_{i}, \cdot)$ is never called twice within a timestep, as discussed in the analysis of time complexity. Moreover, each agent sends a backtracking message at most once in each timestep. Overall, the communication cost for PIBT at each timestep is linear for the number of agents, that is, $O(|A|)$. In reality, this can be even lower because the figures depend on the number of interacting agents $|\mathcal{A}|$, which can be much smaller than $|A|$.

\subsection{Without Rotations}
In practice, in physical environments such as mobile robots, movements corresponding to “rotations” might be difficult to realize due to synchronization problems. A rotation is a set of adjacent agents moving along a circle within one timestep. Formally, a rotation occurs for a subset of agents $\{a_i, a_j, a_k, \ldots, a_l\} \subseteq A$ during timestep $t$ and $t+1$ when $\loc{i}{t+1} = \loc{j}{t} \land \loc{j}{t+1} = \loc{k}{t} \land \ldots \loc{l}{t+1} = \loc{i}{t}$ is satisfied. We refer to a set of paths as \emph{rotation-free} when there is no rotation in the paths. This part adapts PIBT to situations where rotations are prohibited.

PIBT does not require major changes even in a model without rotations. During the iteration for the candidate nodes of the next timestep [Lines~\ref{algo:pibt:func-loop-start}--\ref{algo:pibt:func-loop-end}], skip nodes resulting in rotations like Lines~\ref{algo:pibt:avoid-vertex-conflict} or~\ref{algo:pibt:avoid-swap-conflict} to prevent collisions. We denote the corresponding Algorithm~\ref{algo:pibt} as PIBT$^{\bigotimes}$. This variant outputs a set of rotation-free paths.
\begin{lemma}
  Let $a_1$ denote the agent with highest priority at timestep $t$ and $v^\ast$ the nearest node to $g_1$ neighboring \loc{1}{t}. If $|A| < |V|$ and there exists a path between $v^\ast$ to an arbitrary node without going through $\loc{1}{t}$, PIBT$^{\bigotimes}$ assigns $v^\ast$ to $a_1$ as \loc{i}{t+1}.
  \label{thrm:local-moves-without-rotations}
\end{lemma}
\begin{proof}
  Following $|A| < |V|$, we can derive $\exists v^\prime \not\in \{ \loc{i}{t} \,|\, a_i \in A \}$. There exists a path $\bm{D}$ between $v^{\ast}$ to $v^\prime$ without going through $\loc{1}{t}$. The subsequent proof is almost identical to that of Lemma~\ref{thrm:local-move} and uses $\bm{D}$ instead of $\bm{C}$.
\end{proof}
\begin{theorem}
  Given an MAPF instance such that $G$ is biconnected and $|A| < |V|$, PIBT$^{\bigotimes}$ generates a set of conflict-free and rotation-free paths $\{ \path{1}, \ldots, \path{n} \}$ such that, for any agent $a_i \in A$, $\loc{i}{0} = s_i$ and there is a timestep $t \leq \mathit{diam}(G)\cdot |A|$ such that $\loc{i}{t} = g_i$.
  \label{thrm:reachability-without-rotation}
\end{theorem}
\begin{proof}
  The same proof procedure of the reachability (Theorem~\ref{thrm:reachability}) can be applied using Lemma~\ref{thrm:local-moves-without-rotations} instead of Lemma~\ref{thrm:local-move}.
\end{proof}
Note that this theorem is restricted to biconnected graphs compared to the original theorem on reachability.

\section{Evaluation}
\label{sec:eval}
This section evaluates PIBT thoroughly to empirically demonstrate that PIBT is a quick and scalable MAPF algorithm with acceptable solution quality. PIBT is tested in MAPF (Section~\ref{sec:eval-mapf}) and MAPD (Section~\ref{sec:eval-mapd}), followed by a stress test for the number of agents (Section~\ref{sec:eval-stress}) and robot demonstration (Section~\ref{sec:eval-robot}). The simulator was developed in \cpp{}, and the experiments were run on a laptop with Intel Core i9 2.3GHz CPU and 16GB RAM. The code is available at \url{https://kei18.github.io/pibt2}.
We also provide the experimental result of PIBT for MAPF in extremely dense situations in Section~\ref{sec:dense} of the Appendix.

\subsection{Multi-Agent Path Finding (MAPF)}
\label{sec:eval-mapf}

\subsubsection{Setup}

\paragraph{Benchmark}
The MAPF benchmark~\cite{stern2019definition}, which includes a set of four-connected grids and start--goal pairs for agents, was used. Ten grids were first selected with different portfolios, e.g., size, sparseness, and complexity (see Fig.~\ref{fig:result-mapf-1}, \ref{fig:result-mapf-2} in the Appendix). For each grid, 25 ``random scenarios'' were used while increasing the number of agents by ten up to the maximum (1000 agents in most cases). Therefore, identical instances were tried for the solvers in all settings. Note that almost all selected grids did not satisfy the graph condition of reachability. The problem setting follows conventional MAPF, i.e., agents are eventually on their goals simultaneously.

{
  \setlength{\tabcolsep}{0pt}
  \newcommand{\figwidth}{0.245\linewidth}
  \newcommand{\figrow}[4]{
    \begin{minipage}[t]{1\linewidth}
      \centering
      {\fieldname{#1} (#2; #3)}\medskip\\
      \begin{tabular}{cccc}
        \begin{minipage}{\figwidth}
          \centering
          \includegraphics[width=0.78\linewidth]{fig/raw/mapf/map/#1.pdf}
        \end{minipage}
        & \begin{minipage}{\figwidth}
          \includegraphics[width=1.0\linewidth]{fig/raw/mapf/runtime_vs_costs/runtime_vs_sum_of_costs_#1_#4.pdf}
        \end{minipage}
        & \begin{minipage}{\figwidth}
          \includegraphics[width=1.0\linewidth]{fig/raw/mapf/runtime_vs_makespan/runtime_vs_makespan_#1_#4.pdf}
        \end{minipage}
        & \begin{minipage}{\figwidth}
          \includegraphics[width=1.0\linewidth]{fig/raw/mapf/makespan_vs_costs/makespan_vs_sum_of_costs_#1_#4.pdf}
        \end{minipage}
          \medskip\\
        \begin{minipage}{\figwidth}
          \includegraphics[width=1.0\linewidth]{fig/raw/mapf/success-rate/success-rate_#1.pdf}
        \end{minipage}&
        \begin{minipage}{\figwidth}
          \includegraphics[width=1.0\linewidth]{fig/raw/mapf/runtime/runtime_#1.pdf}
        \end{minipage}&
        \begin{minipage}{\figwidth}
          \includegraphics[width=1.0\linewidth]{fig/raw/mapf/sum-of-costs/sum-of-costs_#1.pdf}
        \end{minipage}&
        \begin{minipage}{\figwidth}
          \includegraphics[width=1.0\linewidth]{fig/raw/mapf/makespan/makespan_#1.pdf}
        \end{minipage}
      \end{tabular}
    \end{minipage}\\
    \begin{minipage}{1\linewidth}
      \includegraphics[width=0.95\linewidth,right]{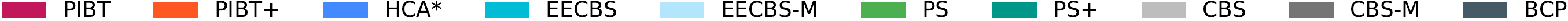}
    \end{minipage}\vspace{0.3cm}\\
  }

  \begin{figure}
    \centering
    \begin{tabular}{l}
      \figrow{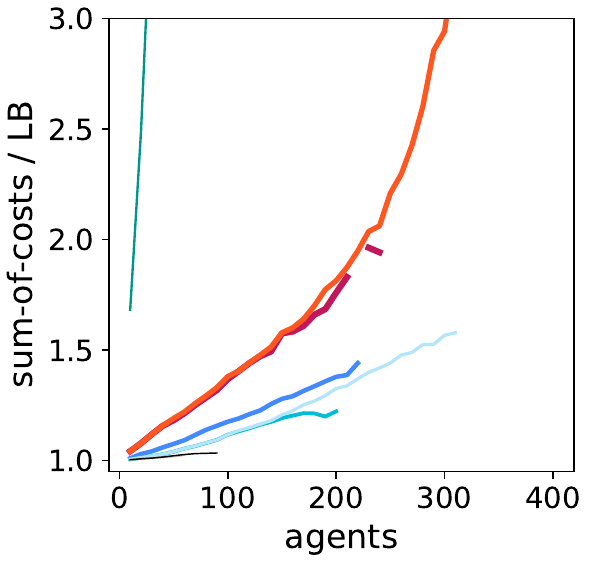}{$32\times 32$}{819}{50}\medskip\\
      \figrow{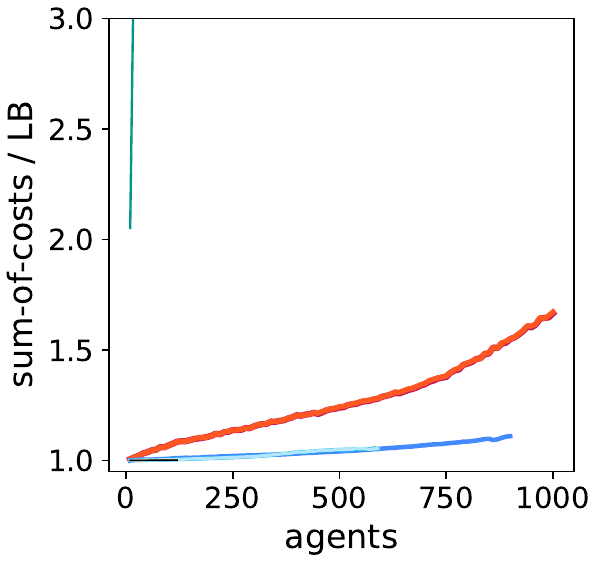}{$194\times 194$}{13,214}{500}
    \end{tabular}
    \caption{
      \textbf{Part of MAPF results.}
      As typical examples, \fieldname{random-32-32-20} and \fieldname{ost003d} were selected.
      The complete results are available in the Appendix.
      In the upper-right part, charts with a fixed number of agents are shown.
      The map size and $|V|$ are shown with parentheses.
      The scores of sum-of-costs for CBS-M and those of makespan for CBS/BCP are omitted.
      In the upper rightmost chart of \fieldname{ost003d}, the plots of EECBS and EECBS-M are overlapped.
    }
    \label{fig:result-mapf-short}
  \end{figure}
}

\paragraph{Baselines}
The following algorithms were carefully picked for comparisons with PIBT in MAPF:
\begin{itemize}
  \item \textbf{\hca}~\cite{silver2005cooperative} as standard prioritized planning~\cite{erdmann1987multiple}.
    Aiming at improving solution quality, we prioritized each agent so that those with more considerable distances from their starts to goals have higher priorities, following heuristics from~\cite{van2005prioritized}.
  \item \textbf{EECBS}~\cite{li2021eecbs} as a state-of-the-art bounded search-based sub-optimal solver; the sub-optimality was set to $1.2$. Note that EECBS is bounded sub-optimal for the sum-of-costs metric.
  \item \textbf{EECBS-M}: The adapted version of EECBS for the makespan metric.
  \item \textbf{Push and Swap (PS)}~\cite{luna2011push} as a standard rule-based approach: although PS is incomplete as pointed out in~\cite{de2013push}, it is the origin of many extended algorithms~\cite{sajid2012multi,de2013push,wiktor2014decentralized,zhang2016discof}; hence, PS was used.
  \item \textbf{PS$^+$}: The post-processing solutions were obtained by PS. PS allows at most one agent to move, resulting in terrible outcomes. Thus, the solutions were compressed while preserving temporal dependencies of the solutions, influenced by the techniques in~\cite{honig2016multi}, which allow multiple agents to move simultaneously.
  \item \textbf{CBS}~\cite{sharon2015conflict} with improvement techniques~\cite{boyrasky2015don,boyarski2015icbs,felner2018adding,li2019improved,li2019disjoint,zhang2020multi,li2021pairwise} as a state-of-the-art search-based optimal solver. Note that CBS optimizes the sum-of-costs metric.
  \item \textbf{CBS-M}: The makespan optimal version of CBS instead of sum-of-costs.
  \item \textbf{BCP}~\cite{lam2019branch,lam2020new} as a state-of-the-art compiling-based optimal solver, which is optimal for the sum-of-costs metric.
\end{itemize}

For EECBS, CBS, and BCP, the implementations coded by their respective authors were used.%
\footnote{
  The codes are available on
  \url{https://github.com/Jiaoyang-Li/EECBS},
  \url{https://github.com/Jiaoyang-Li/CBSH2-RTC}, and
  \url{https://github.com/ed-lam/bcp-mapf}, respectively.
  (EE)CBS-M was implemented in a best-first search manner that prioritizes makespan-better search nodes in the high-level tree, modifying the aforementioned codes.
}
When these implementations have parameter specifications, the default was used for each. The other solvers (\hca, PS, and PS$^+$) were coded by the author, which is own-coded in \cpp{}.

\paragraph{Failure Case}
An algorithm was regarded as having failed to solve an instance when either of the three conditions was met:
(1)~The algorithm reported failure.
(2)~The algorithm reached the runtime limit (30 seconds). This value was based on~\cite{stern2019definition}.
(3)~The algorithm reached the makespan limit (2000 in \fieldname{brc202d}; otherwise 1000). This rules out impractical MAPF outcomes. The values were set according to preliminary results.

\paragraph{Metrics}
The following four evaluation metrics were used, referring to other MAPF studies:
(1)~\textbf{success rate} over 25~instances,
(2)~\textbf{runtime},
(3)~\textbf{sum-of-costs} divided by $\sum_{a_i \in A} \dist{s_i}{g_i}$, and
(4)~\textbf{makespan} divided by $\max_{a_i \in A} \dist{s_i}{g_i}$.
The latter two assess the solution quality; smaller solutions are better, and a minimum of one solution is necessary. Although optimal costs are challenging to obtain, these scores work as a lower bound of sub-optimality. Note that CBS and BCP are optimal for the sum-of-costs metric whereas CBS-M is optimal for makespan; we hence omit to plot makespan scores of CBS/BCP and sum-of-costs scores of CBS-M. Sum-of-costs and makespan have Pareto optimal structure~\cite{yu2013structure}; hence, generally, they cannot be minimized simultaneously.

\paragraph{Other Remarks}
PIBT$^+$ uses PS$^+$ as a complement solver. Own-coded solvers (PIBT$^{(+)}$, \hca, PS$^{(+)}$) used a distance table for the start locations of each agent, computed by breadth-first search. The runtime included the procedure of creating the table.

{
  \setlength{\tabcolsep}{2mm}
  \begin{table}
    \caption{
      \textbf{Classification of failures.}
      The numbers of failed instances, regardless of $|A|$, are reported.
    }
    \label{table:failure}
    \centering
    \fieldname{random-32-32-20}\medskip\\
    {
      \scriptsize
      \begin{tabular}{lrrrrrrrrrr}
        \toprule
        & PIBT & PIBT$^+$ & \hca & EECBS & EECBS-M & PS & PS$^+$ & CBS & CBS-M & BCP
        \\\midrule
        total & 674 & 102 & 590 & 568 & 333 & 908 & 426 & 872 & 862 & 823 \\
        1. failure report & 0 & 3 & 590 & 0 & 0 & 0 & 0 & 0 & 0 & 0 \\
        2. makespan limit & 674 & 99 & 0 & 0 & 0 & 908 & 426 & 0 & 0 & 0\\
        3. time over & 0 & 0 & 0 & 568 & 333 & 0 & 0 & 872 & 862 & 823
        \\\bottomrule
      \end{tabular}
    }
    \vspace{0.5cm}\\
    \fieldname{ost003d}\medskip\\
    {
      \scriptsize
      \begin{tabular}{lrrrrrrrrrr}
        \toprule
        & PIBT & PIBT$^+$ & \hca & EECBS & EECBS-M & PS & PS$^+$ & CBS & CBS-M & BCP
        \\\midrule
        total & 204 & 0 & 473 & 1301 & 1296 & 2497 & 2439 & 2311 & 2237 & 2294 \\
        1. failure report & 0 & 0 & 0 & 0 & 0 & 0 & 0 & 0 & 0 & 0 \\
        2. makespan limit & 204 & 0 & 0 & 0 & 0 & 2497 & 2439 & 0 & 0 & 0 \\
        3. time over & 0 & 0 & 473 & 1301 & 1296 & 0 & 0 & 2311 & 2237 & 2294
        \\\bottomrule
      \end{tabular}
    }
  \end{table}
}

\subsubsection{Result}
Figure~\ref{fig:result-mapf-short} presents part of the results. The full version is displayed in Fig.~\ref{fig:result-mapf-1} and \ref{fig:result-mapf-2} in the Appendix. Excluding success rates, the scores are averaged over only the successful instances with that solver.%
\footnote{
For a fair comparison, it is better to present the average scores of instances that all solvers solved. However, due to the high failure rates, the current style of data plots was selected.
}
In addition, we show the classification of failures of each solver in Table~\ref{table:failure}.

Overall, PIBT outputs solutions immediately ($\leq$ 5 seconds in large maps) even with a thousand agents, with a slight decline in the solution quality (sum-of-costs). Specific findings are summarized as follows:

\paragraph{PIBT is extremely scalable}
The scalability here is defined by map size and the number of agents. In both aspects, PIBT outperforms the other baselines. The runtime of PIBT is significantly faster by orders of magnitude compared to the other solvers. In other words, PIBT can solve large instances that other solvers cannot solve. This is owing to the small-time complexity.

\paragraph{PIBT outputs solutions with acceptable quality in sparse situations}
In general, PIBT does not guarantee solution quality. However, both sum-of-costs and makespan are adequate compared to others in sparse settings. This is in contrast to pure rule-based solvers (PS$^{(+)}$) that mostly fail due to the makespan limit even with fewer agents. PIBT is based on prioritized planning, enabling output solutions with acceptable quality.

\paragraph{Failure reasons of PIBT}
Table~\ref{table:failure} reveals that the failure reason for PIBT is due to the makespan limit. We further explain two failure categories of PIBT as follows. The first case is due to the nature of reachability (Theorem~\ref{thrm:reachability}) that does not ensure that all agents reach their goals simultaneously. The second case is due to graphs without a cycle for any two adjacent nodes. In such graphs, it is possible to reach situations where several agents stop moving, similar to Fig.~\ref{fig:failure}. As those agents remain in their locations, PIBT eventually reaches the makespan limit. Therefore, PIBT often fails with many agents in small maps (for instance, \fieldname{empty-48-48} and \fieldname{random-64-64-20}).

\paragraph{PIBT$^+$ can increase the success rate of PIBT dramatically}
The aforementioned shortcoming of PIBT for one-shot MAPF is overcome by adding a complement solver. Note that PIBT$^+$ sometimes failed because an incomplete solver PS$^+$ was used as a complement solver. Note further that it is possible to use other solvers for the complement solver.

\paragraph{Solution quality of PIBT$^+$ in dense situations}
In Fig.~\ref{fig:result-mapf-short}, the upper bound of suboptimality of PIBT$^+$ in \fieldname{random-32-32-20} dramatically increases with the number of agents. This is due to the complement solver. In the experiment, PIBT$^+$ used PS$^+$ as the complement solver to achieve high success rates within a small computation time. In general, PS$^+$ is quick even for large instances, however, its solution quality is not excellent, as seen in our results. In cluttered and dense situations such as \fieldname{random-32-32-20}, it is difficult to obtain solutions by PIBT. Then, PIBT$^+$ often uses the complement solver, degrading the solution quality.

\paragraph{Interpretations of comparisons' results}
Optimal solvers (CBS and BCP) can address a few hundred agents in some cases, but usually, they cannot address several hundreds of agents, emphasizing the necessity of sub-optimal solvers. Pure rule-based solvers (PS$^{(+)}$) are quick, but their solution qualities are not acceptable compared to the other solvers. Indeed, failures of PS$^{(+)}$ are mainly due to the makespan limit. Prioritized planning (\hca) outputs plausible solutions for both sum-of-costs and makespan. It is scalable but not so much as PIBT. In addition, when scenarios are dense, it is significantly difficult to find solutions according to \emph{static} priorities (see \fieldname{random-32-32-20} in Table~\ref{table:failure}), making \emph{dynamic} priorities attractive as used in PIBT. The overall result of the search-based sub-optimal solver (EECBS) is similar to \hca, i.e., it outputs plausible solutions but is not as quick and scalable as PIBT. In short, PIBT has a unique position compared to other established MAPF approaches, i.e., it is rapid and scalable with acceptable solution qualities.

\subsection{Multi-Agent Pickup and Delivery (MAPD)}
\label{sec:eval-mapd}

\subsubsection{Setup}
The experimental setup follows that of the original MAPD study~\cite{ma2017lifelong}, using the same undirected graph as a testbed, and setting the same locations as candidates for pickup and delivery (Fig.~\ref{fig:mapd-map}).
A sequence of 500~tasks was generated by randomly choosing pickup and delivery locations from all task endpoints. The six different task frequencies where numbers of tasks are added to the task set $\Gamma$ were as follows: 0.2 (one task every five timesteps), 0.5, 1, 2, 5, and 10 with the number of agents increasing from 10 to~50. The Token Passing (TP) algorithm~\cite{ma2017lifelong} was also tested as a baseline, which we programmed in \cpp{}. All experimental settings were performed over 100 instances in which initial positions of agents were set randomly.
The following three metrics were evaluated:
(1)~\textbf{runtime} per one timestep; MAPD is assumed to be online,
(2)~\textbf{service time}: from issue to completion of tasks, and
(3)~\textbf{makespan}: the first timestep at which all tasks are completed.
Both PIBT (Algorithm~\ref{algo:pibt-mapd}) and TP used an all-pairs distance matrix, pre-computed with the Floyd--Warshal algorithm~\cite{floyd1962algorithm}.

{
  \begin{figure}
    \centering
    \includegraphics[width=0.4\linewidth]{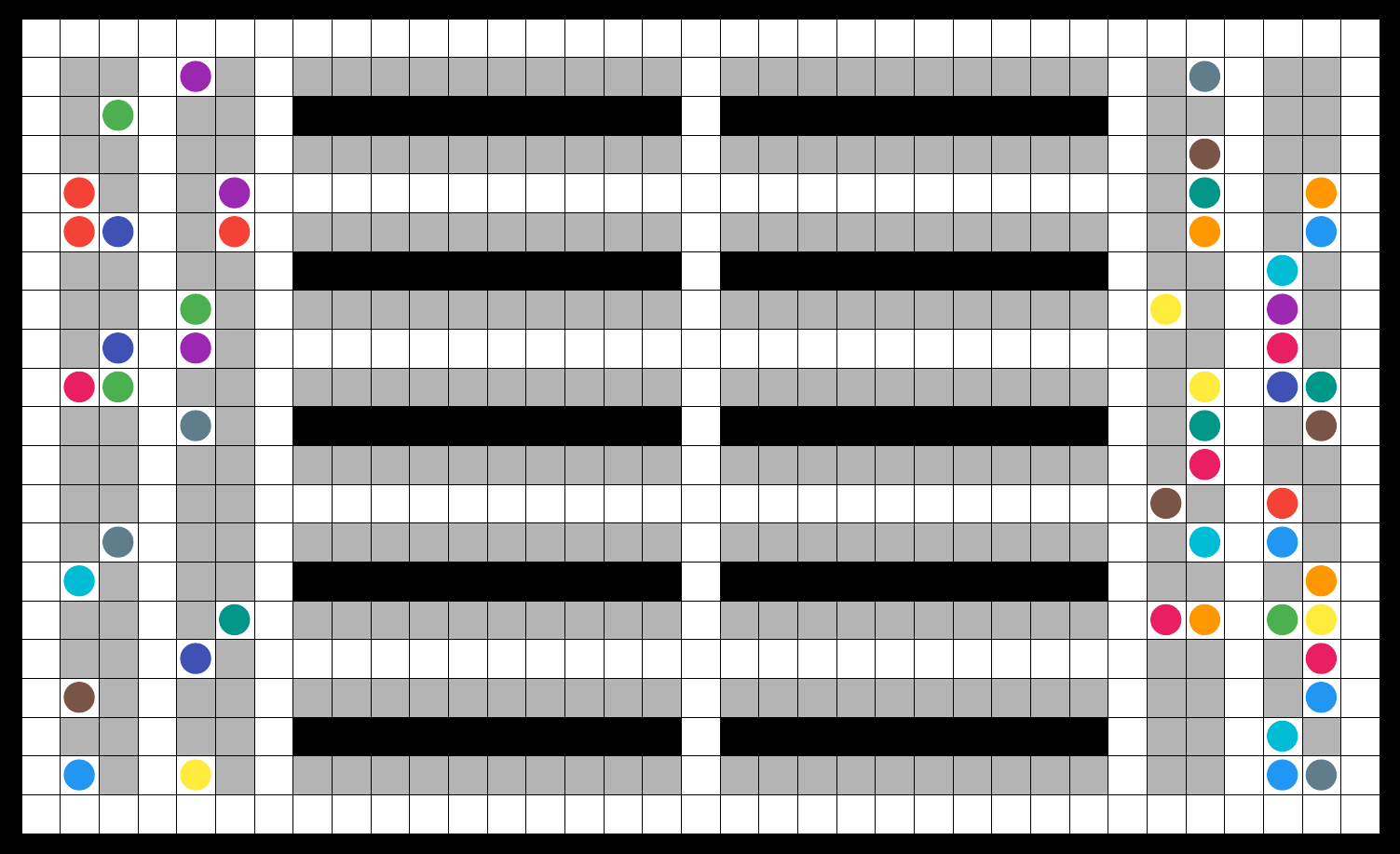}
    \caption{\textbf{Map used in the MAPD experiment.}
      The graph is $21\times 35$ 4-connected grid.
      Gray cells in the MAPD map are task endpoints, i.e., pickup and delivery locations.
    }
    \label{fig:mapd-map}
  \end{figure}
}
{
  \setlength{\tabcolsep}{1pt}
  \newcommand{\imgwidth}{0.85\linewidth}
  \newcommand{\figrow}[1]{
    \begin{minipage}{1.0\linewidth}
      \centering
      \begin{tabular}{cccc}
        \begin{minipage}{0.2\linewidth}
          \center{freq:#1}
        \end{minipage}
        & \begin{minipage}{0.25\linewidth}
          \includegraphics[width=\imgwidth]{fig/raw/mapd/runtime/mapd-runtime_#1.pdf}
        \end{minipage}
        & \begin{minipage}{0.25\linewidth}
          \includegraphics[width=\imgwidth]{fig/raw/mapd/service-time/mapd-service-time_#1.pdf}
        \end{minipage}
        & \begin{minipage}{0.25\linewidth}
          \includegraphics[width=\imgwidth]{fig/raw/mapd/makespan/mapd-makespan_#1.pdf}
        \end{minipage}
      \end{tabular}
    \end{minipage}
  }

  \begin{figure}
    \centering
    \begin{tabular}{c}
      \begin{minipage}{1.0\linewidth}
        \centering
        \begin{tabular}{cccc}
          \begin{minipage}{0.2\linewidth}~\end{minipage}
          & \begin{minipage}{0.25\linewidth}\center{runtime}\end{minipage}
          & \begin{minipage}{0.25\linewidth}\center{service time}\end{minipage}
          & \begin{minipage}{0.25\linewidth}\center{makespan}\end{minipage}
        \end{tabular}
      \end{minipage}\\
      \figrow{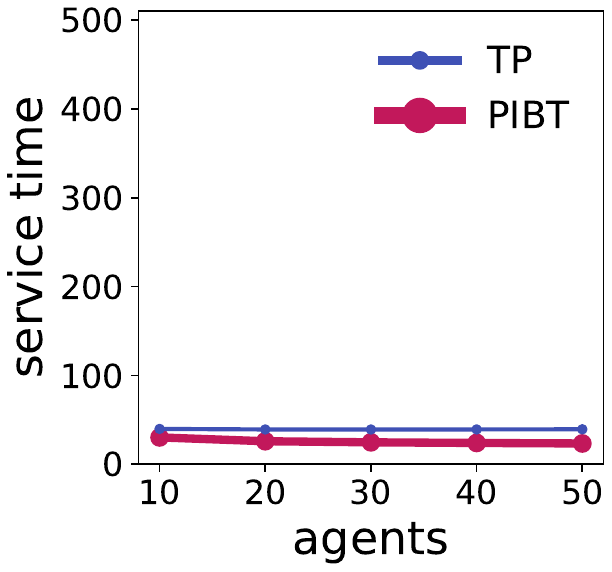}\\
      \figrow{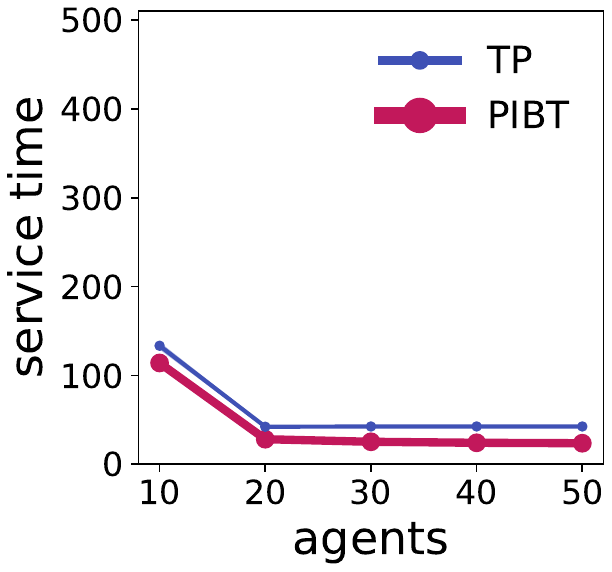}\\
      \figrow{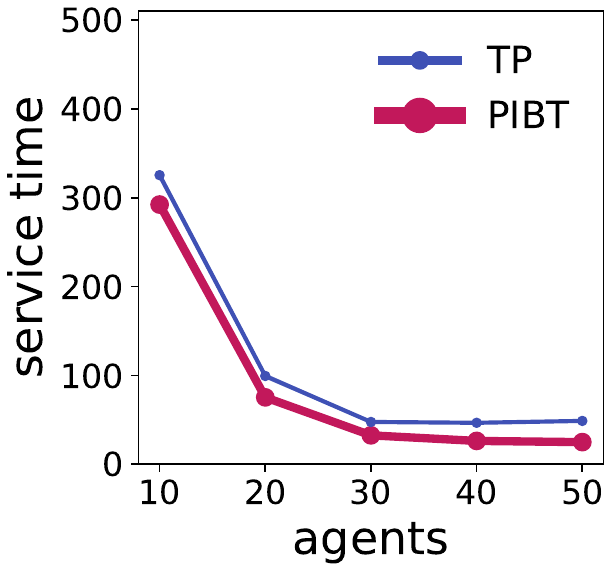}\\
      \figrow{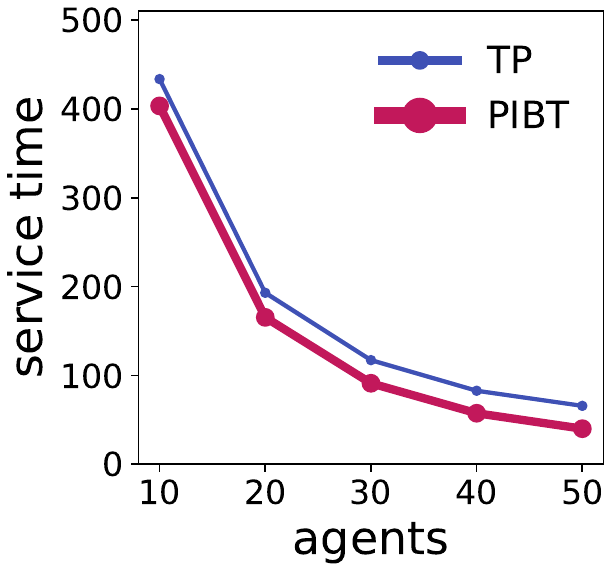}\\
      \figrow{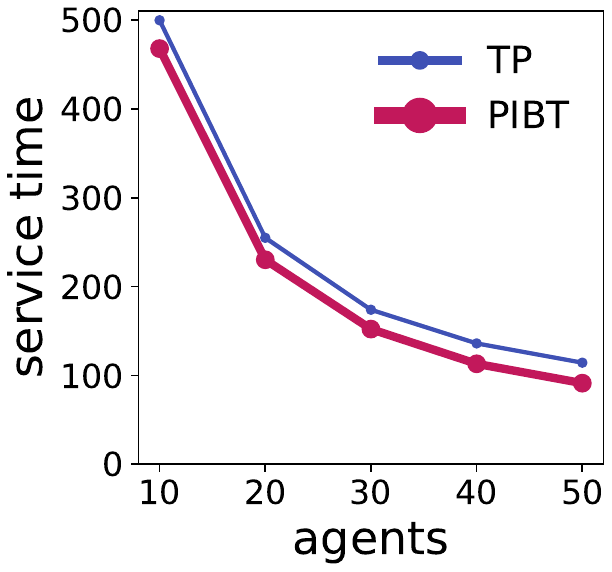}\\
      \figrow{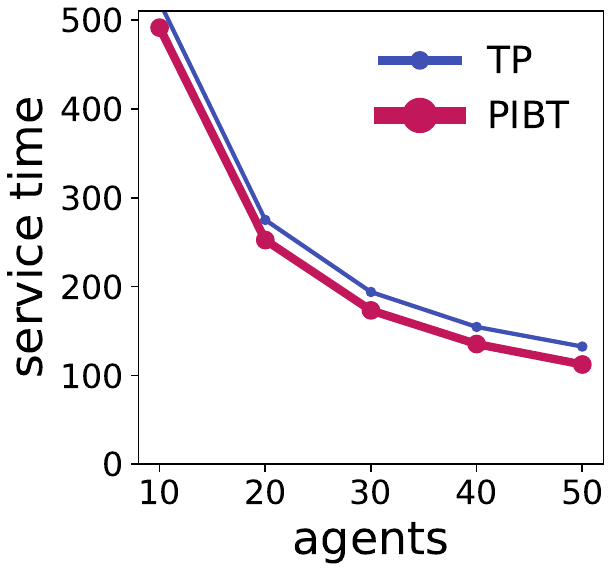}
    \end{tabular}
    \caption{\textbf{Summary of MAPD results.}
      Each value is an average of 100 instances with 95\% confidence intervals.
      For both service time and makespan, the intervals are hard to recognize because they are tiny.
    }
    \label{fig:result-mapd}
  \end{figure}
}

\subsubsection{Result}
Figure~\ref{fig:result-mapd} summarizes the results. PIBT significantly outperforms TP in runtime and is comparable to or better than TP in solution quality, characterized by service time and makespan. The runtime improvements are because of the low time complexity of PIBT. The improvements in solution quality are mainly due to how the algorithm deals with free agents. Assume that there is one unassigned task. In PIBT, all free agents move toward the pickup location, with only the earliest one actually getting the task and then having to start moving to the delivery location while \emph{pushing the other free agents away}, thanks to the prioritization scheme. In contrast, TP evacuates all free agents to non-task endpoints to avoid deadlocks other than the one agent that is newly assigned. However, this assigned agent must still \emph{``dodge'' those free agents' locations} to prevent deadlocks. As a result, the path planned by PIBT for the task-assigned agent will be shorter than that of TP.

\subsection{Stress Test for Scalability}
\label{sec:eval-stress}

{
  \setlength{\tabcolsep}{0pt}
  \begin{figure}
    \centering
    \begin{tabular}{ccc}
      \centering
      \begin{minipage}{0.245\linewidth}
        {\fieldname{orz900d}\vspace{-0.1cm}\\$1491\times 656$ ($96,603$)}
      \end{minipage}
      & \begin{minipage}{0.245\linewidth}
        \centering
        \includegraphics[width=0.78\linewidth]{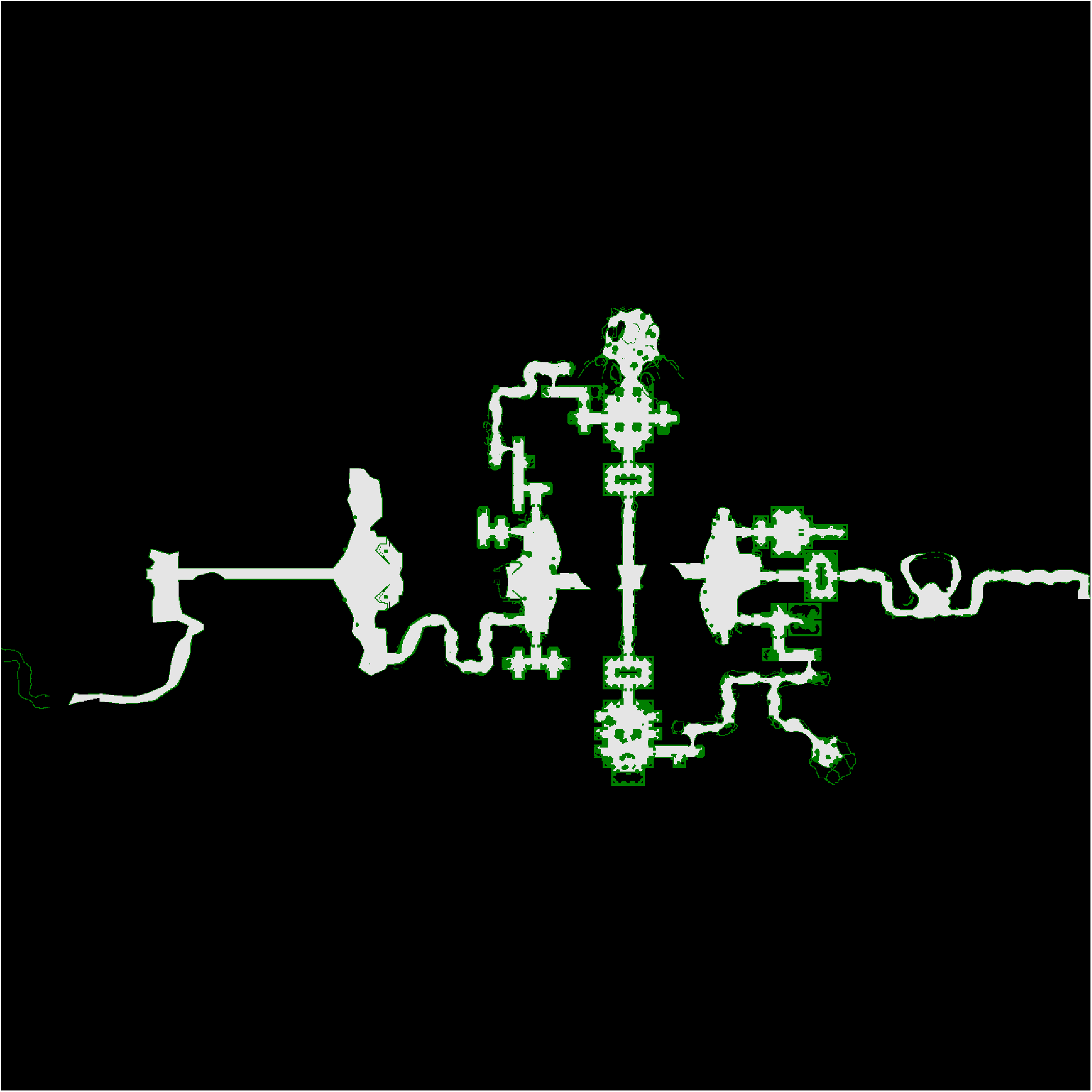}
      \end{minipage}
      & \begin{minipage}{0.5\linewidth}
        \centering
        \includegraphics[width=0.9\linewidth]{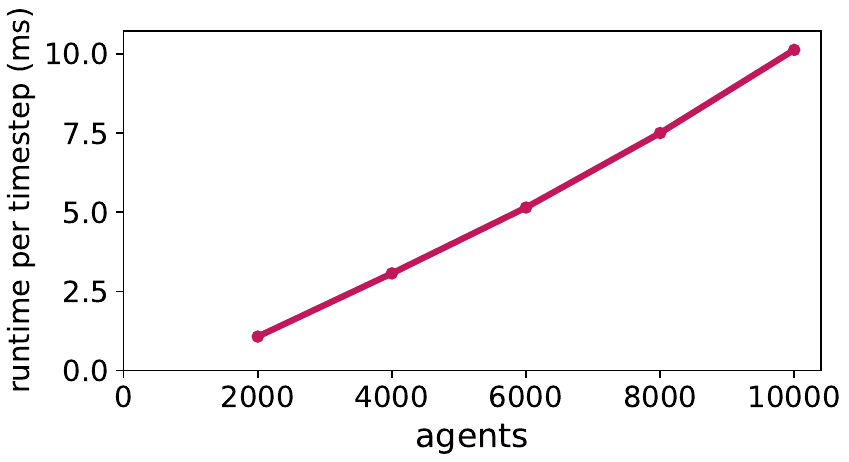}
      \end{minipage}
    \end{tabular}
    \caption{
      \textbf{The results of the stress test.}
      The average runtime per timestep is shown.
      The plots of the 95\% confidence intervals are too small and difficult to recognize.
    }
    \label{fig:result-stress}
  \end{figure}
}

\subsubsection{Setup}
The scalability of PIBT was evaluated through MAPF with a large map while varying the number of agents from $2,000$ to~$10,000$. The map, shown in Fig.~\ref{fig:result-stress}, was the largest in the MAPF benchmark~\cite{stern2019definition}. Here, only runtime was evaluated per timestep, averaged over the first $100$ timesteps, regardless of whether the planning succeeded. Similar to the previous MAPF experiment, distance tables computed before executing PIBT were used. The starts and goals of agents were set randomly for each repetition.

\subsubsection{Result}
Figure~\ref{fig:result-stress} summarizes the result. The scores follow almost a linear trend in the number of agents. Surprisingly, even with $10,000$ agents, PIBT scored around ten milliseconds per one timestep on an ordinary laptop.

\subsection{Demonstrations with Real Robots}
\label{sec:eval-robot}
PIBT was also implemented with a team of small physical robots. Here, an online and lifelong scenario is presented where a new goal is immediately assigned to an agent who reaches its current goal. Note that, because of reachability, PIBT ensures that all assigned goals are eventually met, that is, visited by assigned robots.

\paragraph{Platform}
\emph{Toio} robots (\url{https://toio.io/}) were used to implement PIBT.
The robots, connected to a master computer via the Bluetooth LE (low energy) protocol, advance on a specific playmat and are controllable by instructions using absolute coordinates.

\paragraph{Usage}
The robots were controlled in a \emph{centralized}, \emph{synchronized}, and \emph{online} mode, described as follows.
A virtual grid ($7\times 5$ with obstacles) was created on the playmat; the robots followed the grid. A central server (a laptop) managed the locations of all robots. Periodically, the server executed PIBT for one single timestep and issued the instructions (that is, where to go) to each robot. The code was written in Node.js.

\paragraph{Snapshot}
Figure~\ref{fig:demo} shows a snapshot of the demo. The full video and code are also available at \url{https://kei18.github.io/pibt2}.

{
  \begin{figure}
    \centering
    \includegraphics[width=0.8\linewidth]{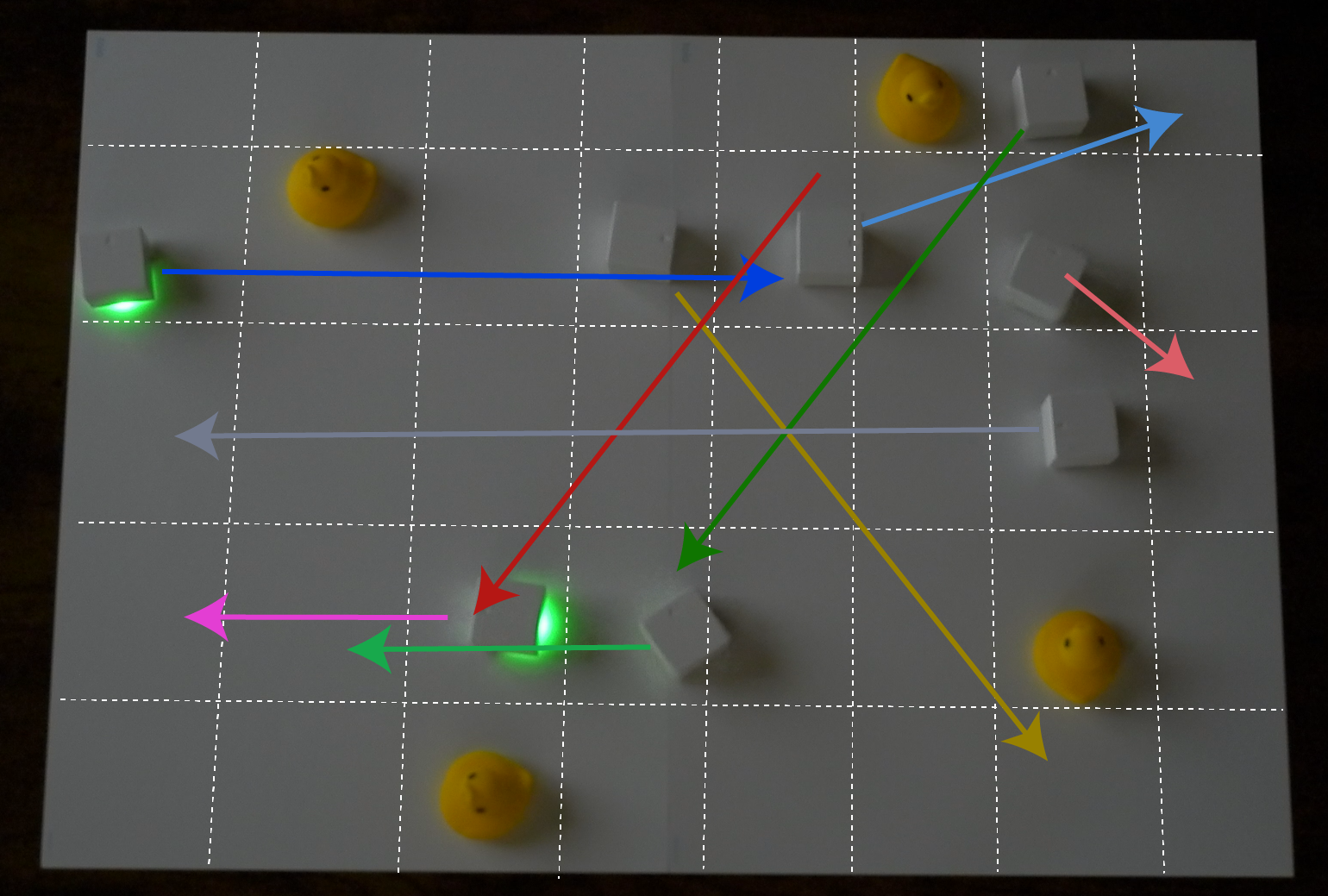}
    \caption{
      \textbf{Snapshot of PIBT demo with real robots.}
      The virtual gird is shown with white dotted lines.
      Each robot's goal is further annotated with colored arrows.
      Green-lighted robots have recently been allocated.
      Yellow ducks are obstacles.
    }
    \label{fig:demo}
  \end{figure}
}

\section{Conclusion}
\label{sec:conclusion}
This study introduces PIBT, a scalable algorithm for solving MAPF iteratively. PIBT focuses on the adjacent movements of multiple agents, relying on a simple prioritization scheme; hence, it can be applied to several domains, including online and lifelong situations. The four empirical results support this aspect:
(1)~PIBT$^{(+)}$ can promptly solve large MAPF instances, whereas other solvers take the time or require unrealistic computational times.
(2)~PIBT outperforms current solutions for pickup and delivery (MAPD).
(3)~Despite thousands of agents, PIBT can yield planning in a real-time manner.
(4)~A real-robot execution of PIBT was further presented.
We believe that PIBT provides practical insights into controlling a swarm of agents in practical situations.

Future directions for research in this field include jumping into the ``wild'' environment, that is, considering \emph{asynchrony}. In traditional MAPF settings, movement between agents is assumed to be synchronized. Although this assumption enables the agents' behavior control, there exists a significant gap in practical situations mainly because of the asynchrony of the agents' moves, which requires further consideration. Our recent study~\cite{okumura2021time} presents other directions.

\section*{Acknowledgments}
We thank the anonymous reviewers for their many insightful comments.
This research is partly supported by JSPS KAKENHI Grant Numbers~20J23011, 21K11748, and 21H03423.
Keisuke Okumura would like to thank the Yoshida Scholarship Foundation for their support.
We thank Editage (\url{www.editage.com}) for English language editing.

\bibliographystyle{unsrt}
\bibliography{ref}
\newpage
\appendix
\section*{Appendix}

\section{MAPF Results}
Figures~\ref{fig:result-mapf-1} and~\ref{fig:result-mapf-2} summarize all results of the MAPF experiments in Section~\ref{sec:eval-mapf}.

{
  \setlength{\tabcolsep}{0pt}
  \newcommand{\figwidth}{0.195\linewidth}
  \newcommand{\figrow}[3]{
    \begin{minipage}[t]{1\linewidth}
      \centering
      \begin{tabular}{ccccc}
        \begin{minipage}{0.24\linewidth}
          \centering
          {\scriptsize \fieldname{#1}\vspace{-0.1cm}\\#2 (#3)\\}
          \includegraphics[width=0.5\linewidth]{fig/raw/mapf/map/#1.pdf}
        \end{minipage}&
        \begin{minipage}{\figwidth}
          \includegraphics[width=1.0\linewidth]{fig/raw/mapf/success-rate/success-rate_#1.pdf}
        \end{minipage}&
        \begin{minipage}{\figwidth}
          \includegraphics[width=1.0\linewidth]{fig/raw/mapf/runtime/runtime_#1.pdf}
        \end{minipage}&
        \begin{minipage}{\figwidth}
          \includegraphics[width=1.0\linewidth]{fig/raw/mapf/sum-of-costs/sum-of-costs_#1.pdf}
        \end{minipage}&
        \begin{minipage}{\figwidth}
          \includegraphics[width=1.0\linewidth]{fig/raw/mapf/makespan/makespan_#1.pdf}
        \end{minipage}
      \end{tabular}
    \end{minipage}\\[5mm]
  }
  \newcommand{\figrowheaders}[4]{
    \begin{minipage}[t]{1\linewidth}
      \centering
      \begin{tabular}{ccccc}
        \begin{minipage}{0.24\linewidth}
          \mbox{~~}
        \end{minipage}&
        \begin{minipage}{\figwidth}
            \quad\hfill{#1}\hfill~
        \end{minipage}&
        \begin{minipage}{\figwidth}
            \quad\hfill{#2}\hfill~
        \end{minipage}&
        \begin{minipage}{\figwidth}
            \quad\hfill{#3}\hfill~
        \end{minipage}&
        \begin{minipage}{\figwidth}
            \quad\hfill{#4}\hfill~
        \end{minipage}
      \end{tabular}
    \end{minipage}\\[2mm]
  }
  \newcommand{\figrowlegend}{
    \begin{minipage}{1\linewidth}
      \includegraphics[width=0.95\linewidth,right]{fig/raw/mapf/labels.pdf}
    \end{minipage}\vspace{0.3cm}\\
  }

  \begin{figure}
    \centering
    \begin{tabular}{l}
      \figrowheaders{Success rate}{Runtime}{Sum-of-costs}{Makespan}
      \figrow{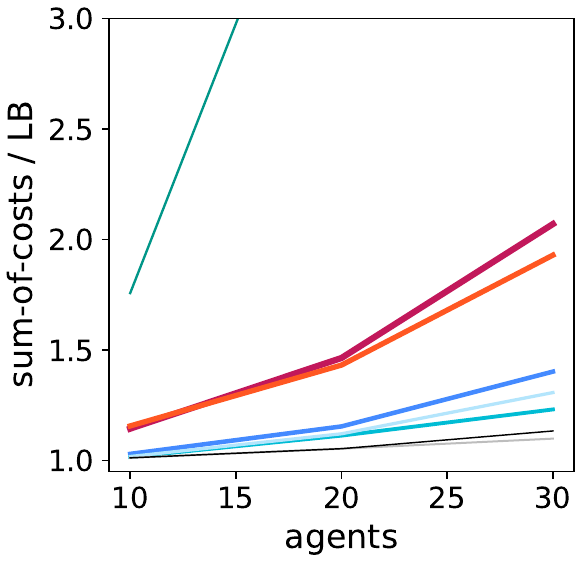}{$8\times 8$}{64}
      \figrow{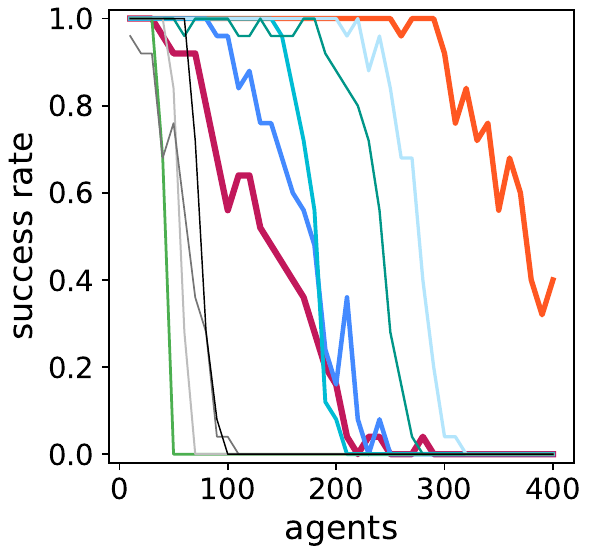}{$32\times 32$}{819}
      \figrow{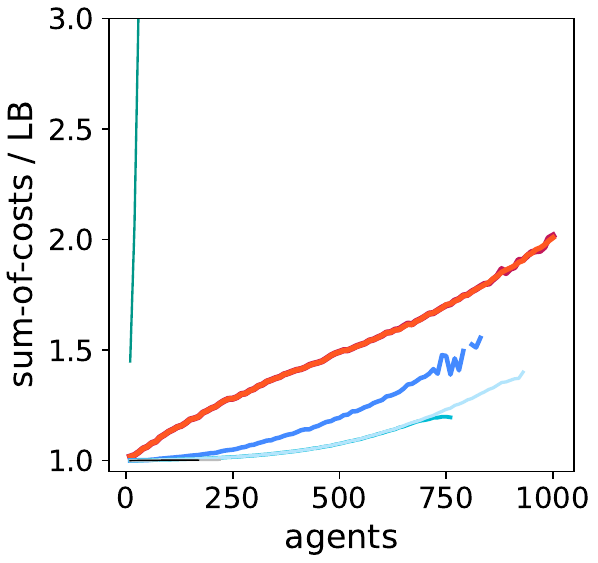}{$48\times 48$}{2,304}
      \figrow{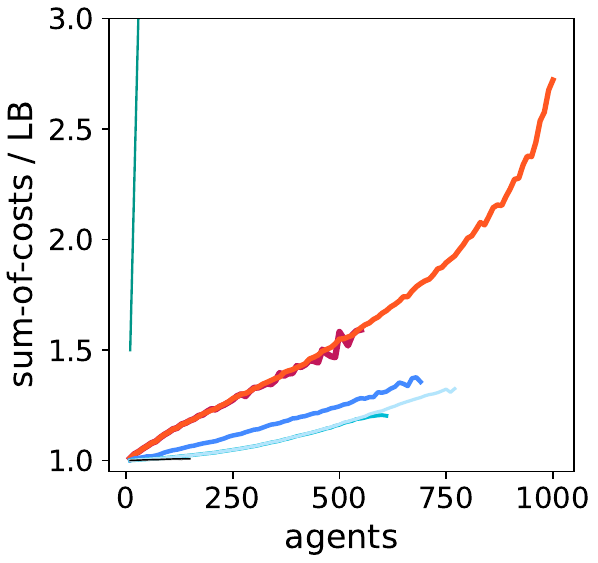}{$64\times 64$}{3,687}
      \figrow{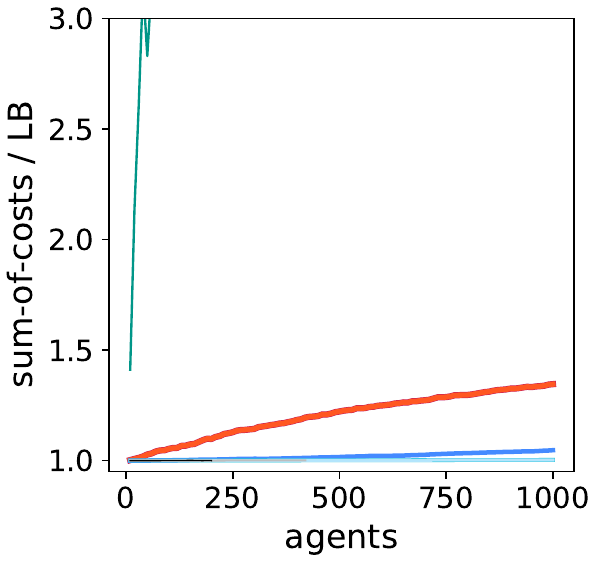}{$340\times 160$}{38,756}
      \figrowlegend
    \end{tabular}
    \caption{
      \textbf{Summary of MAPF results (1/2).}
      $|V|$ is shown in parentheses.
    }
    \label{fig:result-mapf-1}
  \end{figure}

  \begin{figure}
    \centering
    \begin{tabular}{l}
      \figrowheaders{Success rate}{Runtime}{Sum-of-costs}{Makespan}
      \figrow{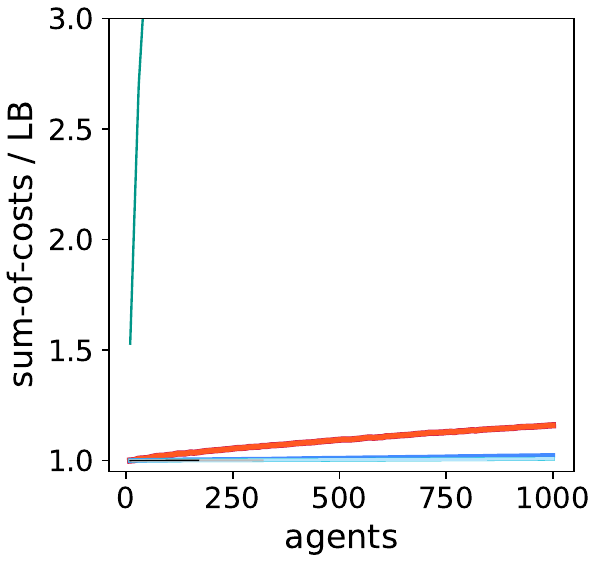}{$256\times 256$}{47,540}
      \figrow{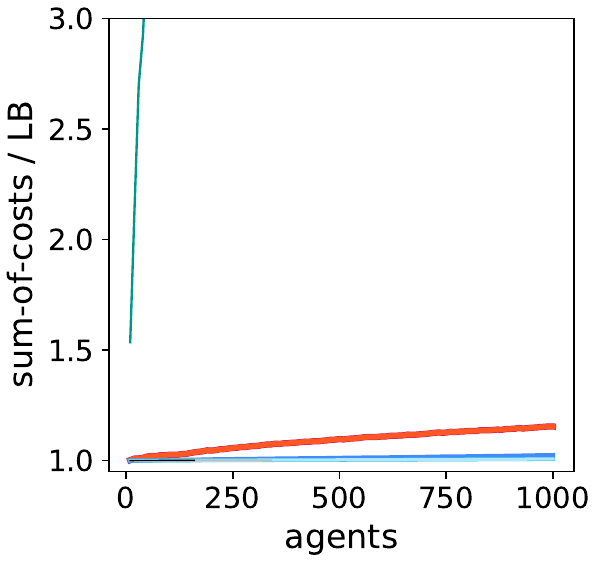}{$256\times 256$}{47,240}
      \figrow{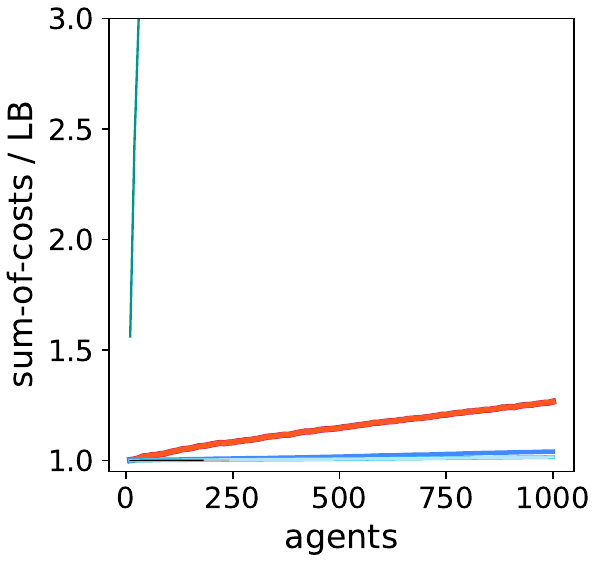}{$256\times 257$}{28,178}
      \figrow{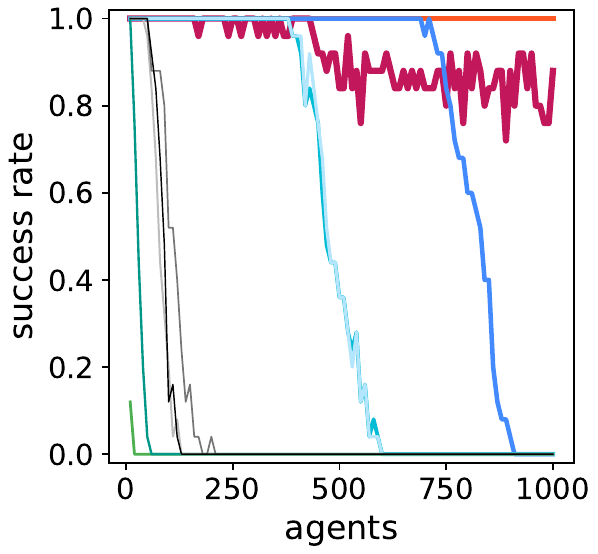}{$194\times 194$}{13,214}
      \figrow{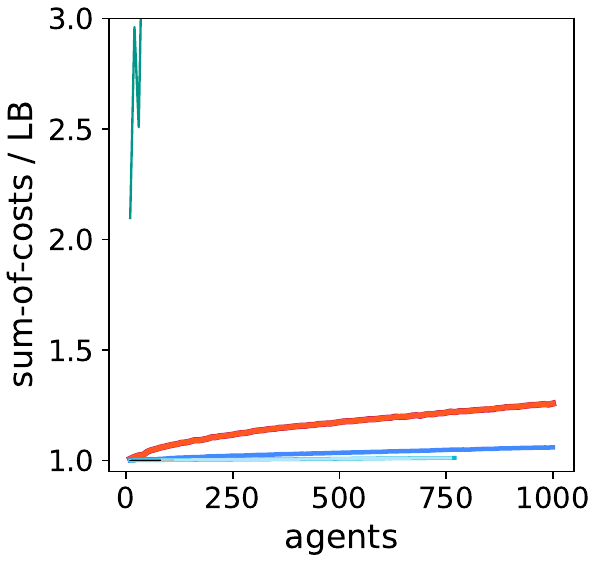}{$530\times 481$}{43,151}
      \figrowlegend
    \end{tabular}
    \caption{
      \textbf{Summary of MAPF results (2/2).}
      $|V|$ is shown in parentheses.
    }
    \label{fig:result-mapf-2}
  \end{figure}
}

\section{PIBT in Extremely Dense Situations}
\label{sec:dense}

This section presents additional experiments on PIBT in extremely dense situations where $|A| > |V|/2$ and $G$ satisfies the graph condition of Theorem~\ref{thrm:reachability}.
We used \fieldname{empty-8-8} and the same experimental settings as those introduced in Section~\ref{sec:eval-mapf}.

Table~\ref{table:extremely-dense} summarizes the result.
Most instances are solved at most 5~ms; otherwise, PIBT failed by reaching the makespan limit.
Solution qualities dramatically increase with more agents because, in dense situations, most agents cannot take their shortest paths.

{
  \begin{table}[ht!]
    \caption{
      \textbf{PIBT in extremely dense situation.}
      The used map is \fieldname{empty-8-8}.
      25 instances were prepared for each $|A|$.
      Scores of ``sum-of-costs'' and ``makespan'' are upper bounds of sub-optimality in the same way as Fig.~\ref{fig:result-mapf-short}.
    }
    \label{table:extremely-dense}
    \centering
    \begin{tabular}{ccccc}
      \toprule
      $|A|$ & success rate & runtime (ms) & sum-of-costs & makespan
      \\\midrule
      40 & 0.96 & 0.21 & 3.15 & 3.46 \\
      50 & 0.84 & 1.43 & 7.38 & 6.94 \\
      60 & 1.00 & 2.16 & 12.25 & 7.86 \\
      64 & 1.00 & 3.16 & 21.55 & 10.01
      \\\bottomrule
    \end{tabular}
  \end{table}
}

Interestingly, PIBT solved all instances with $64$ agents (fully occupied), while sometimes failing with $50$ agents.
This is because of the randomness of the tie-break rule at Line~\ref{algo:pibt:sort-candidates} of Algorithm~\ref{algo:pibt}.
As a discussion of the implementation level, we used three rules when sorting candidate nodes for the following location: (1) distances to the goal, (2)~the presence of agents to avoid unnecessary priority inheritance, and (3)~random values, when neither the prior two break a tie.
We also tested the rule-2 omitted version, succeeding for all instances regardless of $|A|$.
Since the rule-2 loses its meaning in situations where $|A| = |V|$, we consider that the randomness contributes to solving extremely dense situations.
We also observed that this trend is the same when testing a $16\times 16$ empty grid, i.e., usual PIBT sometimes failed whereas PIBT that omits the rule-2 tie-break solved all instances regardless of $|A|$.
From these observations, we guess that in tidy environments like \fieldname{empty-8-8}, PIBT without the rule-2 can solve one-shot MAPF with a high probability as makespan increases.
This is an interesting direction to seek but significantly beyond this paper.

Note that the aforementioned discussion is not applicable when $G$ does not satisfy the graph condition of Theorem~\ref{thrm:reachability}.
Moreover, rule-2 is usually effective in improving sum-of-costs, as shown in Table~\ref{table:solution-quality-appendix}.

{
  \newcommand{\myblock}[1]{\renewcommand{\arraystretch}{0.5}\begin{tabular}{c}#1\end{tabular}}
  \newcommand{\ci}[1]{\scriptsize(#1)}
  \begin{table}[ht!]
    \centering
    \caption{
      \textbf{Effect of tie-break strategy.}
      ``normal'' is PIBT and ``random'' is PIBT without the rule-2 tie-break.
      The used map was \fieldname{den520d}.
      For each $|A|$, 25 random scenarios were prepared, not equivalent to those in Sec.~\ref{sec:eval-mapf}.
      The scores of sum-of-costs and makespan are average of upper bounds of sub-optimality with 95\% confidence intervals, based on instances that were solved by both.
    }
    \label{table:solution-quality-appendix}

    \begin{tabular}{ccccccc}
      \toprule
      & \multicolumn{2}{c}{success rate}
      & \multicolumn{2}{c}{sum-of-costs}
      & \multicolumn{2}{c}{makespan}
      \\\cmidrule(lr){2-3}\cmidrule(lr){4-5}\cmidrule(lr){6-7}
      $|A|$
      & normal & random & normal & random & normal & random
      \\\midrule
      100
      & 1.00
      & 1.00
      & \myblock{1.04\\\ci{1.04, 1.05}}
      & \myblock{1.08\\\ci{1.07, 1.09}}
      & \myblock{1.00\\\ci{1.00, 1.00}}
      & \myblock{1.00\\\ci{1.00, 1.00}}
      \\
      300
      & 1.00
      & 1.00
      & \myblock{1.10\\\ci{1.09, 1.10}}
      & \myblock{1.16\\\ci{1.15, 1.16}}
      & \myblock{1.00\\\ci{1.00, 1.00}}
      & \myblock{1.00\\\ci{1.00, 1.00}}
      \\
      500
      & 0.96
      & 1.00
      & \myblock{1.15\\\ci{1.14, 1.15}}
      & \myblock{1.22\\\ci{1.22, 1.23}}
      & \myblock{1.00\\\ci{1.00, 1.00}}
      & \myblock{1.00\\\ci{1.00, 1.00}}
      \\
      700
      & 0.96
      & 0.96
      & \myblock{1.20\\\ci{1.20, 1.20}}
      & \myblock{1.28\\\ci{1.27, 1.28}}
      & \myblock{1.00\\\ci{1.00, 1.00}}
      & \myblock{1.00\\\ci{1.00, 1.00}}
      \\
      900
      & 0.88
      & 0.96
      & \myblock{1.25\\\ci{1.24, 1.25}}
      & \myblock{1.33\\\ci{1.32, 1.34}}
      & \myblock{1.00\\\ci{1.00, 1.00}}
      & \myblock{1.00\\\ci{1.00, 1.01}}
      \\\bottomrule
    \end{tabular}
  \end{table}
}

\end{document}